\documentclass[aps,pre,amsmath,amssym,amsthm,superscriptaddress,reprint, nofootinbib]{revtex4-2}
\usepackage[colorlinks=true,urlcolor=blue,citecolor=blue,linkcolor=blue]{hyperref}

\usepackage{graphicx}
\usepackage{dcolumn}
\usepackage{kotex}
\usepackage{natbib}
\usepackage{color}
\usepackage{amsfonts}
\usepackage{hyperref}
\usepackage{algorithm}
\usepackage{algcompatible}
\usepackage{newfloat}

\usepackage{makecell}
\usepackage{bm}
\usepackage{setspace}
\usepackage{multirow}
\usepackage{sidecap}
\usepackage{enumerate}

\definecolor{mygray}{gray}{0.5}

\begin{document}
\title{Interplay of network structure and talent configuration on wealth dynamics}

\author{Jaeseok Hur}
\affiliation{Department of Physics, Korea Advanced Institute of Science and Technology, Daejeon 34141, Korea}

\author{Meesoon Ha}
\email[Contact author;]{msha@chosun.ac.kr}
\affiliation{Department of Physics Education, Chosun University, Gwangju 61452, Korea}

\author{Hawoong \surname{Jeong}}
\affiliation{Department of Physics, Korea Advanced Institute of Science and Technology, Daejeon 34141, Korea}
\affiliation{Center of Complex Systems, KAIST, Daejeon 34141, Korea}

\begin{abstract}
The economic success of individuals is often determined by a combination of talent, luck, and assistance from others. We introduce an agent-based model that simultaneously considers talent, luck, and social interaction. This model allows us to explore how network structure (how agents interact) and talent distribution among agents affect the dynamics of capital accumulation through analytical and numerical methods. We identify a phenomenon that we call the ``talent configuration effect," which refers to the influence of how talent is allocated to individuals (nodes) in the network. We analyze this effect through two key properties: talent assortativity (TA) and talent-degree correlation (TD). In particular, we focus on three economic indicators: growth rate ($n_{\rm rate}$), Gini coefficient (inequality: $n_{\rm Gini}$), and meritocratic fairness ($n_{LT}$). This investigation helps us understand the interplay between talent configuration and network structure on capital dynamics. We find that, in the short term, positive correlations exist between TA and TD for all three economic indicators. Furthermore, the dominant factor influencing capital dynamics depends on the network topology. In scale-free networks, TD has a stronger influence on the economic indices than TA. Conversely, in lattice-like networks, TA plays a more significant role. Our findings address that high socioeconomic homophily can create a dilemma between growth and equality and that hub monopolization by a few highly talented agents makes economic growth strongly dependent on their performances.
\end{abstract}

\maketitle

\section{\label{intro}Introduction}

\begin{figure}[t]
\includegraphics[width=1\columnwidth]{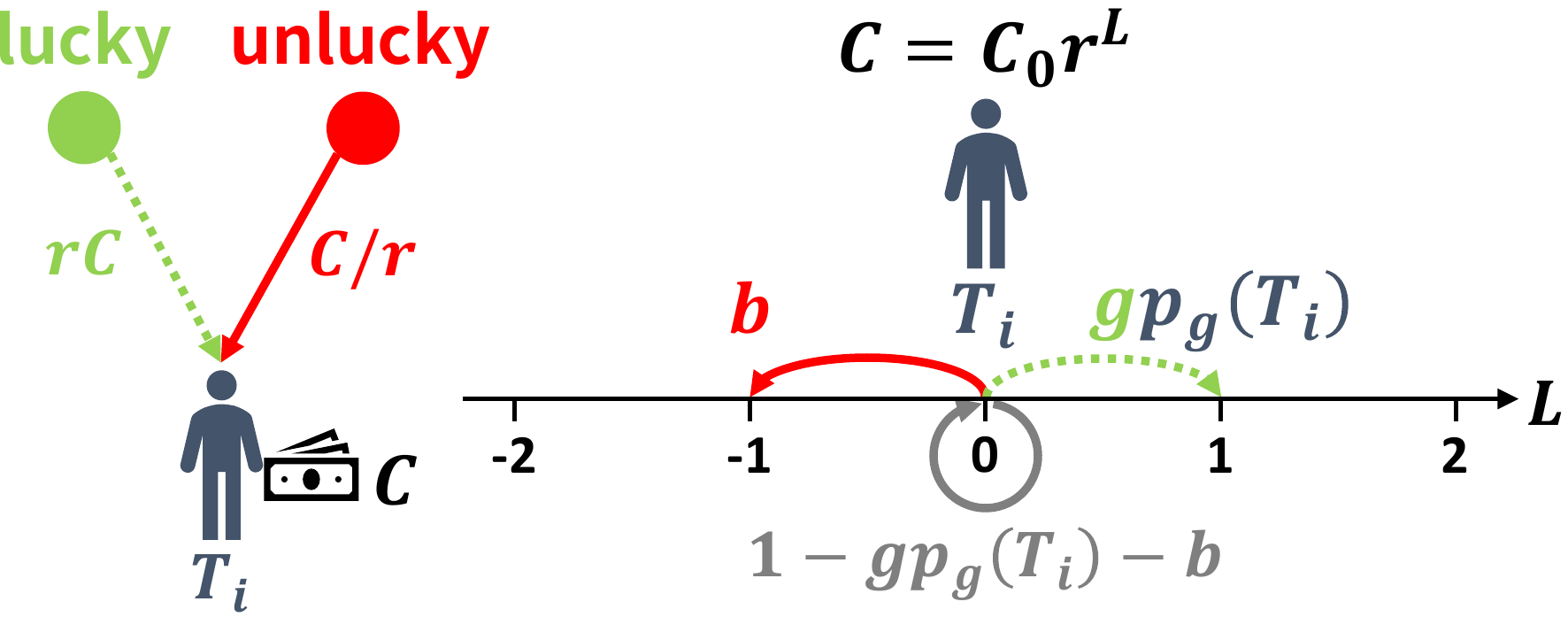}
    \caption{Dynamics of talent versus luck (TvL) model. In the left panel, an agent $i$ with talent $T_i$ and capital $C$ meets either a lucky event with probability $g$ or an unlucky one with probability $b$. The lucky event gives chance to multiply capital by the factor $r>1$, and the unlucky event always divides capital by $r$. In the right panel, the TvL model is illustrated as a one-dimensional random walk with probability $gp_g(T_i)$, $b$, and $1-gp_g(T_i)-b$ to move from the original site in the capital level $L$ space at the right to the left and stay, where $L=m-n$ with the number of lucky (unlucky) events $m$ ($n$).}
    \label{fig1-model}
\end{figure}

It is always questionable which of talent, luck, and innate environment has the greatest impact in an individual's success. Pluchino {\it et al.}~\cite{pluchino2018talent} recently proposed the ``talent versus luck'' (TvL) model (see Fig.~\ref{fig1-model}) to quantitatively assess the impact of talent and luck on an individual's success. In the TvL model, the number of good and bad events represents the total amount of opportunities for either the positive or negative aspects of the environment. They showed that under a mediocre environment with the same number of good and bad events, an individual's talent is not strongly related to success, whereas under a good environment with more good events than bad events, high talent tends to guarantee more success. It implies the importance of the environment, in which an individual's talent can be fully realized. In addition to the total opportunities, the agents with which an individual interacts can also play a crucial role in wealth dynamics, further emphasizing environmental effects. Barab{\'a}si emphasized the importance of networks in the universal laws of success~\cite{barabasi2018formula}, and Zhou {\it et al.}~\cite{zhou2023nature} also presented a generative model for network growth, in which nature (fitness) and nurture (social advantage) effects act simultaneously.

In this paper, we propose a general framework for capital dynamics in agent-based networks. We introduce a model incorporating talent, luck, and social interaction (TLS). In the TLS model, talent acts as a fitness factor, increasing average capital accumulation. Luck introduces random fluctuations in capital holdings, while social interaction directs capital transfers towards more connections during interagent exchange. Consequently, we analytically demonstrate that the TLS model can reproduce the earlier results of the TvL model~\cite{pluchino2018talent} and the Bouchaud-M{\'e}zard model (BM) model~\cite{bouchaud2000wealth}, because the TvL model and the BM model correspond to the TLS model without social interactions and agent talent heterogeneity, respectively.

Unlike the BM model and its extensions~\cite{souma2001small, souma2003wealth, garlaschelli2004wealth, garlaschelli2008effects}, where capital dynamics solely depend on network structure, in the TLS model, we consider both network structure and the distribution of talent across the network (talent configuration). This allows us to explore the impact of the talent configuration on interactions among heterogeneous agents for a given network. We analyze this effect through three key economic indicators: growth rate, Gini coefficient (inequality), and meritocratic fairness.

In order to quantify talent configuration, we employ two key properties: talent assortativity (TA) and talent-degree correlation (TD). Our findings show a positive correlation between both TA and TD with three metrics in the short-time regime. In addition, the dominant factor influencing these indices depends on the network topology. In scale-free networks, TD has a stronger impact compared to TA, whereas the opposite holds true for lattice-like networks.

The remainder of this paper is organized as follows: In Sec.~\ref{model}, we propose a model of wealth dynamics, which covers two earlier models, the TvL model and the BM model. In Sec.~\ref{results}, we define two talent configuration (TC) properties in the TLS model to speculate how TC properties affect three economic indicators: growth rate, Gini coefficient, and meritocratic fairness. In Sec.~\ref{summary}, we conclude by summarizing our findings with some remarks. Detailed mathematical derivations and explanations are also provided as well as mean-field calculations of the TLS model in Appendixes.

\section{\label{model} Model}

We propose a model of wealth dynamics with talent, luck, and social interaction (TLS), which covers both the ``talent versus luck" (TvL) model~\cite{pluchino2018talent} and the Bouchaud-M{\'e}zard (BM)~\cite{bouchaud2000wealth} model, named the TLS model. In the TLS model, agent $i~(1\le i\le N)$ has talent $T_i$, degree $k_i$, and time-dependent capital $C_i(t)$ at time $t$. The last one is extrinsic, whereas the first two are intrinsic.

In order to study the capital change of $N$ agents in the TLS model, similar to the TvL model with social interaction among neighboring agents (see Appendix~\ref{TvL} for the detailed analysis of the original TvL model), we denote the capital set, ${\bf C}(t)=\{C_1(t), C_2(t), \ldots, C_N(t)\}$, the talent set, ${\bf T}=\{T_1,T_2,\ldots,T_N\}$ following a normal distribution $T_i\sim \mathcal{N}(\mu, \sigma^2)$ with the mean talent $\mu$ and the standard deviation $\sigma$, and the degree set, ${\bf k}=\{k_1, k_2,\ldots,k_N\}$. Initially, every agent starts with the same capital as $C_i(0)=C_0$.

As illustrated in Fig.~\ref{fig1-model}, the capital of each agent, $C$, can be updated by the following parameters: the capital multiplier $r(>1)$, and the event probabilities of lucky, unlucky, and nothing-happened events, $\{gp_g(T),b,1-gp_g(T)-b\}$, respectively, where $g+b\le 1$ and $p_g(T)=\min\{1,\max[T,0]\}$. The probability of winning capital for the lucky event, $gp_g(T)$, implies that for the agent with talent $T$, the probability to gain the capital $rC$ is proportional to the probability of a lucky event occurring and the probability that $\mbox{rand}[0,1]$ falls below $T$. In the discrete time update from $t$ to $t+1$, one of three events occurs to every agent: (1) For a lucky event, a random number is generated from a uniform distribution between 0 and 1 (written as $\mbox{rand}[0,1]$). If $\mbox{rand}[0,1]<T_i$, agent $i$'s capital is increased by $r$ times, $C_i(t+1)=rC_i(t)$. (2) For an unlucky event, its capital is reduced by $1/r$ times: $C_i(t+1)=C_i(t)/r$. (3) When nothing has happened, its capital remains the same as before, $C_i(t+1)=C_i(t)$.

Since $r$ is a constant, the amount of capital per agent is determined by the number of times for the agent to win and lose the capital by rules (1) and (2). Let $m$ and $n$ represent the number of times the agent wins and loses capital according to rules (1) and (2), respectively ($m=0,1,...$, $n=0,1,...$), so that the capital level $L\equiv m-n$. Then the amount of capital is equal to $C=C_0r^L$. The change of $L$ at each time step, $\Delta L$, can have a value among $\{-1, 0, 1\}$. However, the probability per each agent differs by the talent of agent. It can be considered as the ensemble of random walkers~\cite{simao2021talent} (see Fig.~\ref{fig1-model}). For the agent with talent $T$, $\Delta L$ satisfies the following probability mass function:
\begin{align}
    p_{L}(\Delta L)=
    \begin{cases}
    b & \text{for}\ \Delta L=-1, \\
    1-gp_g(T)-b & \text{for}\ \Delta L=0,\\
    gp_g(T) & \text{for}\ \Delta L=+1.\\
    \end{cases}
    \label{eq-P_L}
\end{align}

In the original TvL model, capital discontinuously changes at each time step. To construct a continuous model, we employ the well-known geometric Brownian motion (GBM) (see Appendixes~\ref{TvL} and ~\ref{SDE-TvL} for detailed mathematical derivations and additional explanations from the discrete version of the TvL model to the continuous version), then we can write the stochastic differential equation (SDE) for the capital of agent $i$, $C_i(t)$, such that
\begin{align}
dC_i(t)=\alpha(T_i)C_i(t)dt+\beta(T_i)C_i(t)dW_{t,i}
    \label{eq-TvL}
\end{align}

This is the continuous version of the TvL model, where $dt$ is time interval, $W_{t,i}$ is the Wiener process of agent $i$ at time $t$, $\alpha(T_i)$ is the percentage drift as a function of talent $T_i$, and $\beta(T_i)$ is the percentage volatility, respectively. The detailed definitions of $\alpha$ and $\beta$ by the identity of $C$ and $L$ can be found in Appendix~\ref{SDE-TvL}. In addition, by introducing the social interaction between agents, we denote the SDE for the TLS model as follows:
\begin{align}
    dC_i(t) = \alpha(T_i)C_i(t)dt+\beta(T_i) C_i(t)dW_{t,i}\nonumber\\
    +\sum_{j(\neq i)}\left[J_{ij}C_j(t)-J_{ji}C_i(t)\right]dt.
    \label{eq-TLS-prev}
\end{align}

The last interaction term in Eq.~\eqref{eq-TLS-prev} describes the capital transfer by exchange among neighboring agent pairs $(i,j)$, which is also suggested in the BM model~\cite{bouchaud2000wealth}. Summing up together for all $i$, the interaction term cancels out to 0. This implies that exchange itself does not change the total capital at a given time. However, since all agents have different $\alpha$ and $\beta$, capital transfers between heterogeneous agents can promote or disrupt average capital growth of a system eventually.

\begin{figure}[t]
\includegraphics[width=0.9\columnwidth]{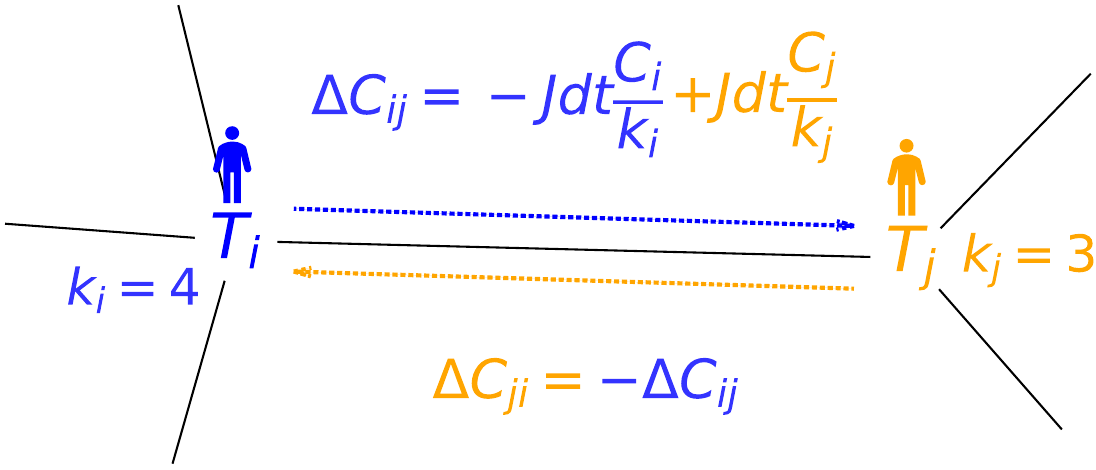}
    \caption{Illustration of pooling sharing interaction~\cite{stojkoski2019cooperation}. The capital of $JdtC_i/k_i$ is given to all neighbors of agent $i$, and agent $j$ does in the same manner~\cite{PS}.
    }
    \label{fig2-PS}
\end{figure}

In this paper, we introduce the matrix element $J_{ij}$ as the pooling and sharing~\cite{stojkoski2019cooperation} interaction, so that every agent acts as both a sharing node and a pooling node as
\begin{align}
  J_{ij}=
    \begin{cases}
    J/k_j & \text{if}\ a_{ij}=1, \\
    0 & \text{if}\ a_{ij}=0, \\
    \end{cases}
    \label{eq-J}
\end{align}
where $a_{ij}$ is the adjacency matrix element for a given network, either 1 or 0, $J(>0)$ is the exchange strength, and $k_j$ is the degree of agent $j$; see Fig.~\ref{fig2-PS}. 

For the simplicity, Eq.~\eqref{eq-TLS-prev} can be rewritten with Eq.~\eqref{eq-J}, $\alpha_i=\alpha(T_i)$, and  $\beta_i=\beta(T_i)$ as follows:
\begin{align}
    dC_i(t) =\ & \alpha_iC_i(t)dt+\beta_iC_i(t)dW_{t,i}\nonumber\\
    & -Jdt\sum_{\{j|a_{ij}=1\}}\left(\frac{C_i(t)}{k_i}-\frac{C_j(t)}{k_j}\right).
    \label{eq-TLS}
\end{align}

As shown in Fig.~\ref{fig1-model} and Fig.~\ref{fig2-PS}, capital dynamics in the TLS model is influenced by individual talent, luck, and interaction among neighboring agents. The interaction term of Eq.~\eqref{eq-TLS} represents capital transfer between the connected link of agent $i$ and $j$, where agent $i$ gives $JdtC_i(t)/k_i$ the amount of capital to $j$, and $j$ does vice versa.

For the case of $C_i(t)=C_j(t)$, the agent with the larger degree gains the more capital from the opponent by exchange. We call this kind of advantage ``high-degree advantage." Hence, in the TLS model, higher talent and higher degree are advantageous to capital growth. It can be regarded as individual advantage (or nature) and social advantage (or nurture) for each agent, in the context of the concept suggested by Zhou {\it et al.}~\cite{zhou2023nature}.

Moreover, in the TLS model, the network structure described by $a_{ij}$ also affects capital dynamics. For the complete network described by adjacency matrix $a_{ij}=1-\delta_{ij}$, where $\delta_{ij}$ is Kronecker delta, 1 if $i=j$ and 0 otherwise. For this case, the interaction term simply becomes as
\begin{align}
    -Jdt\left[\left(\frac{N}{N-1}\right)C_i-\left(\frac{N}{N-1}\right)\bar{C}\right]\simeq -Jdt(C_i-\bar{C}),
    \label{eq-J_approx}
\end{align}
where $N$ is large enough. The term of $-Jdt(C_i-\bar{C})$ is exactly the same as the mean-field interaction suggested by Bouchaud and M{\'e}zard~\cite{bouchaud2000wealth}. Therefore, the TLS model represented by Eq.~\eqref{eq-TLS} on a large complete network is approximately the same as the mean-field TLS model (see Appendix~\ref{MF-TLS} for analytic results of the mean-field TLS model).

Before moving on to Sec.~\ref{results}, we would like to make a couple of interesting remarks concerning mean-field models: (1) The BM model without interaction is exactly the same as the GBM. The mean capital of both the GBM and the mean-field BM model are the same and exponential. (2) The TLS model without interaction is exactly the same as the TvL model. However, the mean capital of the TvL model is not exponential, whereas the mean-field TLS model is. Hence, it is found that the mean-field interaction plays such a different role in the GBM and the TvL model.

In addition, regarding the power-law tail exponent $\gamma$, talent heterogeneity does not change $\gamma$ of noninteractive models but changes $\gamma$ for mean-field models. The characteristics of each wealth dynamics model is summarized in Table~\ref{table:1} (see the details in Appendix~\ref{MF-TLS}).

\section{\label{results} Talent Configuration (TC) Effect}

In order to analyze the effect of talent configuration (TC) properties on capital dynamics, we consider the TLS model on agent-based networks (see Fig.~\ref{fig2-PS}). The most interesting point is that in the TLS model, different talent allocations on a given network give different capital dynamics. We call this the TC effect. For node index $i(1\leq i\leq N)$, the talent configuration is defined as a vector $\bold{T} = (T_1,\dots,T_N)$.

Due to the fact that the talent distribution follows a normal distribution, $T_i\sim\mathcal{N}(\mu,\sigma^2)$, there are a huge number of cases that allocate talent samples to nodes on a given network, but we cannot investigate all of those cases. Therefore, it is important to extract statistical properties of talent configuration that consistently correlate with capital dynamics in the TLS model. These will be discussed in Sec.~\ref{TA+TD}. Similarly, capital dynamics is also influenced by the four environmental parameters $(r,g,b,J)$ and network structure $a_{ij}$. It is noted that $(r, g,b,J)$ and $a_{ij}$ correspond to capital multiplier, lucky and unlucky event probabilities, exchange strength, and the elements of adjacency matrix, respectively.

In this study, to consider the TC effect only, we set the environmental parameters as $(r,g,b,J)=(2.0,0.1,0.1,0.1)$ unless described, and analyze two representative network cases, the Barab{\'a}si-Albert (BA) network~\cite{barabasi1999emergence} with the degree heterogeneity and the scale-free property, and the Watts-Strogatz (WS) network~\cite{watts1998collective} with the small rewiring probability $p_{\rm re}=0.1$ as well as the small-world property.

For the BA network generation, the linear preferential attachment is used with an additional link attachment per node. More precisely, the mean degree $\bar{k}$ becomes $2(1-1/N)$, so that $\bar{k}\to 2$ for $N\gg 1$. For the WS network generation, $\bar{k}=2$ and the rewiring probability $p_{\rm re}=0.1$ are chosen. When $p_{\rm re}=0$, the WS network is equal to a cycle network. When $p_{\rm re}=1$, all links are randomly rewired to others. Our choice of the parameter $p_{\rm re}=0.1$ is small enough for one to consider that the WS network is more close to a lattice-like network.

The reason why we use $\bar{k}=2$ for those network cases is rather simple. If $\bar{k}$ is very small, a random network generates many isolated nodes. Since the TvL model corresponds to capital dynamics of all isolated nodes and has no TC effect, the smaller $\bar{k}$ gives the TvL-like capital dynamics and the less TC effect. Meanwhile, if $\bar{k}$ is very large, the interaction term of Eq.~\eqref{eq-TLS} becomes close to Eq.~\eqref{eq-J_approx}. Since the mean-field interaction corresponds to an all-to-all connection and has no TC effect, the more $\bar{k}$ gives the mean-field TLS-like capital dynamics and the less TC effect. Therefore, to avoid the generation of isolated nodes and to have a lower network density as possible for both BA and WS networks, we set $\bar{k}=2$.

As the estimators of the TC effect, we focus on three economic indices: the growth rate ($n_{\rm rate}$), the Gini coefficient ($n_{\rm Gini}$), and the meritocratic fairness ($n_{LT}$), with the following definitions:
\begin{align}
    n_{\rm rate}\equiv\frac{\langle C\rangle}{C_0}, 
    \label{eq-growth}
\end{align}
which is the index for the growth rate that represents how many times the system has grown, $n_{\rm rate}\in[0,\infty]$.

\begin{align}
    n_{\rm Gini} & \equiv\frac{1}{2\langle C\rangle}\int_{-\infty}^{\infty}\int_{-\infty}^{\infty}p(C)p(C')|C-C'|dCdC', 
    \label{eq-Gini}
\end{align}
which is the index for the inequality (Gini coefficient) that represents the inequality depicted by a Lorenz curve, $n_{\rm Gini}\in[0,1]$. As a result, $n_{\rm Gini}=0$ means system is perfectly equal and $n_{\rm Gini}=1$ means system is perfectly unequal,

\begin{align}
    n_{LT} & \equiv\frac{\langle LT\rangle-\langle L\rangle \langle T\rangle}{\sqrt{Var(L)Var(T)}},
    \label{eq-LT}
\end{align}
which is the index for the meritocratic fairness that  represents how much talent and following reward are related to each other, defined as Pearson correlation coefficient between the capital level $L=\log_r(C/C_0)$ and talent $T$, $n_{LT}\in[-1,1]$. So $n_{LT}=-1$ means that $L$ and $T$ are perfectly anti-correlated (meritocratically unfair), $n_{LT}=0$ means that there is no correlations between $L$ and $T$ (meritocratically neutral), and $n_{LT}=1$ means that $L$ and $T$ are perfectly correlated (meritocratically fair).

\begin{figure}[]
\includegraphics[width=\columnwidth]{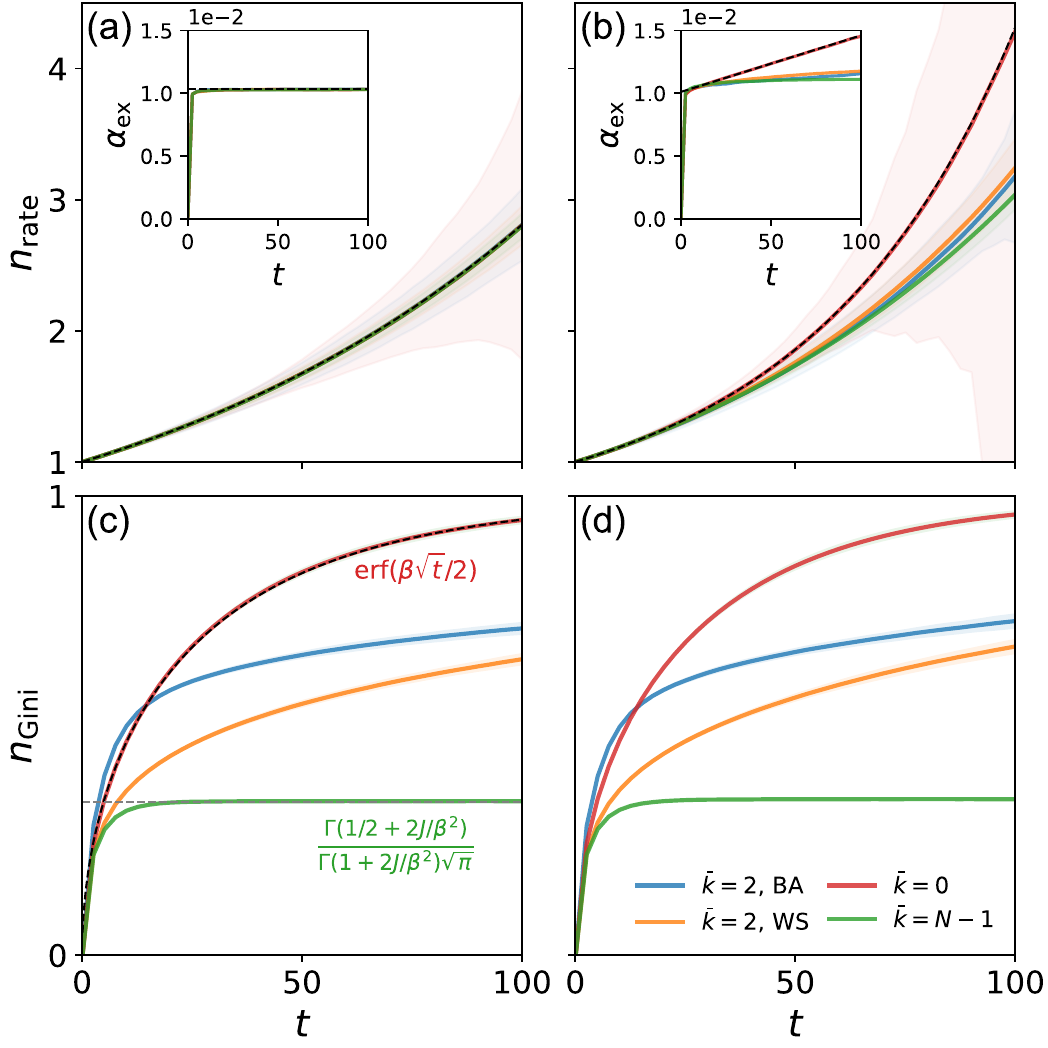}
    \caption{Heterogeneous talent effect on growth rate and Gini coefficient: $n_{\rm rate}$ for (a) and (b), and $n_{\rm Gini}$ for (c) and (d) against time $t$, where $\bar{k}=\{0, 2({\rm BA}), 2({\rm WS}), N-1\}$. Numerical results for a fixed talent $T_i=0.6$ [(a) and (c), left] are compared with those for Gaussian talents $T_i\sim\mathcal{N}(0.6,0.1^2)$ [(b) and (d), right]. All solid lines (colored regions) show the average (the standard deviation) of $2^{10}$ realizations, except for red solid lines (colored regions) with $2^{15}$ realizations. Black and gray dashed lines are the analytical baselines for $\bar{k}=0$ and $\bar{k}=N-1$ cases, respectively. For (a) and (b), two insets show the time evolution of $\alpha_{\rm ex}$ denoted in Eq.~\eqref{eq-alpha_ex}. It is noted that for all cases, $r_{\rm TA}=0$, $r_{\rm TD}=0$, or they are not defined; see more details in Eq.~\eqref{eq-r_TA} and Eq.~\eqref{eq-r_TD} of Sec.~\ref{TA+TD}. All simulation results are obtained for the same parameters $(N,r,g,b,J)=(10^4,2,0.1,0.1,0.1)$ unless described.} 
    \label{fig3-T_i}
\end{figure}

\subsection{\label{hetero-T} Heterogeneous Talent effects on Networks: Growth rate and Inequality}

For the analysis of capital dynamics between different talent configurations, in this subsection, we first investigate how different talent distributions, Dirac-delta distribution of talents (fixed) and the normal distribution of talents (Gaussian), affect capital dynamics in the TLS model, respectively. Particularly to estimate such effects against the talent heterogeneity, we control the environment, network topology, and talent configurations. 

In Fig.~\ref{fig3-T_i} we plot the time evolution of $n_{\rm rate}$ and $n_{\rm Gini}$ as a function of time for the very sparse random network case ($\bar{k}=2$) of the BA network and the WS network ($p_{\rm re}=0.1$), which are guided by two limiting cases: the non-network case ($\bar{k}=0$,no interaction among agents) and the complete network case ($\bar{k}=N-1$, all-to-all). It is noted that all talent samples are randomly allocated for those cases, so that there are no correlations between talent configuration and network structure (i.e., $r_{\rm TA}=0$ and $r_{\rm TD}=0$; see Sec.~\ref{TA+TD}).

\begin{figure*}[]
\includegraphics[width=\textwidth]{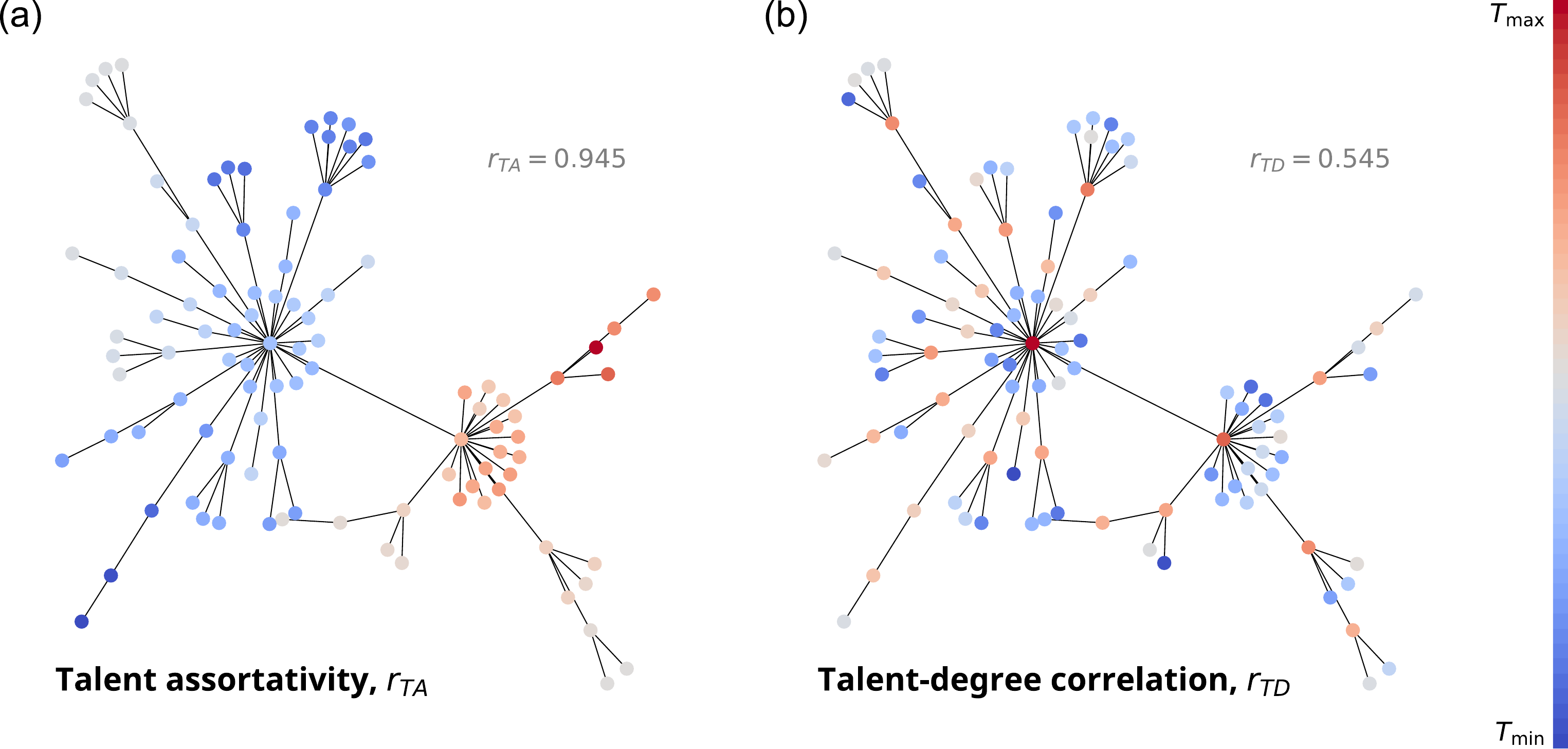}
    \caption{Conceptual visualization for two talent configuration properties: (a) TA ($r_{\rm TA}$) and (b) TD ($r_{\rm TD}$). The sample network is the BA network with the number of agents $N=100$ and the number of links $M=99$. Nodal colors indicate high (red) or low (blue) talents, and talent samples follow the normal distribution $T\sim\mathcal{N}(\mu,\sigma^2)$. It is noted that $T_{\rm max}$ ($T_{\rm min}$) is the maximum (minimum) value for given talent samples. (a) An example for the high TA case, where nodes with similar talents are clustered; (b) an example for high TD case, where highly talented nodes are allocated to hubs for a given network.} 
    \label{fig4-TA_TD}
\end{figure*}

In the TLS model, the $\bar{k}=0$ case with fixed talent corresponds to the GBM, and that with Gaussian talent corresponds to the TvL model. Similarly, the $\bar{k}=N-1$ case with fixed talent corresponds to the mean-field BM model, and that with Gaussian talent corresponds to the mean-field TLS model.

In the top panel of Fig.~\ref{fig3-T_i}, Figs.~\ref{fig3-T_i}(a) and~\ref{fig3-T_i}(b) show the time evolution of $n_{\rm rate}$, where all curves are the same for the cases with fixed talent but different only for those with Gaussian talent. It implies that the system growth does not depend on the network structure with fixed talent, whereas on that with Gaussian talent, it does. This result becomes a key difference between the BM model and the TLS model. We also argue the growth effective network structure or talent configuration in the TLS model, unlike the BM model. Insets show the time evolution of $\alpha_{\rm ex}$, defined in Eq.~\eqref{eq-alpha_ex}. For fixed talent, all $\alpha_{\rm ex}$ values converge to the same value and grow exponentially, whereas for Gaussian talent, they become different and grow super-exponentially in the short-time regime.

In the bottom panel of Fig.~\ref{fig3-T_i}, Figs.~\ref{fig3-T_i}(c) and~\ref{fig3-T_i}(d) show the time evolution of the inequality $n_{\rm Gini}$, where they mainly depend on network structure, but there are only slight differences between the fixed talent and Gaussian talent cases. The fixed talent cases with $\bar{k}=0$ and $\bar{k}=N-1$ are given as
\begin{align}
    n_{\rm Gini}(t) &= {\rm erf}(\beta\sqrt{t}/2) && \text{for $\bar{k}=0$},
    \label{eq-fixed_T+n_Gini-1} \\
    \lim_{t\rightarrow\infty}n_{\rm Gini}(t) & = \frac{\Gamma(1/2+2J/\beta^2)}{\Gamma(1+2J/\beta^2)\sqrt{\pi}} && \text{for $\bar{k}=N-1$},
    \label{eq-fixed_T+n_Gini-2}
\end{align}
which are the smallest and the largest mean degree. Both cases become the solvable baselines of $n_{\rm Gini}$. However, for $\bar{k}=2$ in the intermediate region, $n_{\rm Gini}$ depends on the network structure. In the short-time regime, the BA network has larger $n_{\rm Gini}$ than that of the WS network for any case, either fixed or Gaussian talent.

\subsection{\label{TA+TD} Talent Configuration (TC) properties:\\ Talent Assortativity and Talent-Degree correlation}

In this subsection we quantify talent configuration properties on capital dynamics as two measures: ``talent assortativity" (TA), $r_{\rm TA}$, and ``talent-degree correlation" (TD), $r_{\rm TD}$, as illustrated in Fig.~\ref{fig4-TA_TD}. 

By the link-based analysis, the TA property is denoted as
\begin{align}
    r_{\rm TA}\equiv\frac{Cov(T,T')}{\sqrt{Var(T)Var(T')}}=\frac{\sum_{T}\sum_{T'}TT'(e_{TT'}-q_{T}q_{T'})}{\sum_{T}T^2q_T-(\sum_{T}Tq_T)^2}.
    \label{eq-r_TA}
\end{align}
The $r_{\rm TA}$ value is Pearson correlation coefficient for all links' talents $T$ and $T'$ for a given network, where $e_{TT'}$ is the probability that talent $T$ and $T'$ are connected in the network, $q_T$ is the probability that the nodes of randomly selected link have talent as $T$. By the node-based analysis, the TD property is denoted as 
\begin{align}
    r_{\rm TD}\equiv\frac{Cov(T,k)}{\sqrt{Var(T)Var(k)}}.
    \label{eq-r_TD}
\end{align}
The $r_{\rm TD}$ value is Pearson correlation coefficient for all nodes' talent $T$ and degree $k$.

A conceptual visualization for TA and TD properties are shown in Fig~\ref{fig4-TA_TD}, where both $r_{\rm TA}$ and $r_{\rm TD}$ are increased as possible by pair swapping algorithms, see SM, pseudo-codes in Table S1~\cite{SM}. While $r_{\rm TA}$ reflects how many similar talents are connected for a given network, $r_{\rm TD}$ reflects how much higher talents tend to have higher degrees for a given network. If the elements of the talent vector $\bold{T}$ are randomly permutated, these two talent configuration properties become almost 0.

For given a network and talent samples, we use a pair swapping algorithm to find a talent configuration $\bold{T}'$ that has a specific value of $r_{\rm TA}$ (or $r_{\rm TD}$) as we wish. The pair swapping algorithm for $r_{\rm TA}$ control consists of five steps:
\begin{enumerate}[(1)]
    \item Start with a graph $G$ and a talent vector $\bold{T}$.
    \item Select a random pair of nodes for the graph $G$.
    \item Let $\tilde{\bold{T}}$ be a talent vector, where selected two nodes are switched. For a target value $r'$, if $|r_{\rm TA}(G,\tilde{\bold{T}})-r'|<|r_{\rm TA}(G,\bold{T})-r'|$, accept $\tilde{\bold{T}}$ as a new $\bold{T}$.
    \item Repeat 2 to 3 unless $|r_{\rm TA}-r'|<\epsilon$.
    \item If $|r_{\rm TA}-r'|<\epsilon$, stop and print $\bold{T}$.
\end{enumerate}

By the same algorithm, we also control $r_{\rm TD}$. However, this algorithm may not guarantee a target value $r'$. Although both $r_{\rm TA}$ and $r_{\rm TD}$ are defined as Pearson correlation coefficients and they lie in the interval $[-1, 1]$, they do not mean that the minimum and maximum of $r_{\rm TA}$ and $r_{\rm TD}$ are $-1$ and $1$, respectively. 

Actual bounds for $r_{\rm TA}$ and $r_{\rm TD}$ depend on the detail of network structures. If we set a target value $r'$ to out of the real bound and set a error range $\epsilon$ to sufficiently a small value, the while loop in step 4 never ends. It is noted that the TA bounds were studied in~\cite{cinelli2020network}. Unlike TA bounds, TD bounds are easier to be controlled because they depend on the node-based analysis that does not influenced by the complex network topology. If all nodes have different degrees for a network, the minimum and maximum values of $r_{\rm TD}$ are exactly equal to $-1$ and $1$, respectively, which are characterized by degree heterogeneity and were studied in~\cite{jacob2017measure}.

For a given network, both $r_{\rm TA}$ and $r_{\rm TD}$ are measured. To test the pure $r_{\rm TA}$ ($r_{\rm TD}$) effect, one prefers to fix $r_{\rm TD}$ ($r_{\rm TA}$) as 0. However, both $r_{\rm TA}$ and $r_{\rm TD}$ are correlated under the single pair swapping. Therefore, we need to control two talent configuration properties at once in the random pair swapping algorithm. The pseudo-codes of algorithms for the dual value control are also summarized in SM as Table S1~\cite{SM} with some illustrations. The core argument of estimating a single talent configuration property is that the other one is in the sufficiently small error range of $\epsilon$, and we consider that this pair of($r_{\rm TA}$, $r_{\rm TD}$) gives the quasi-pure effect of $r_{\rm TA}$ (or $r_{\rm TD}$). Throughout this procedure, we estimate the quasi-pure effect of $r_{\rm TA}$ (or $r_{\rm TD}$) on capital dynamics in the TLS model (see Fig.~\ref{fig5-BA+WS-TA+TD}).

\subsection{\label{T+short_time} Short-term Behaviors of TC effects}

In this subsection, we present how to investigate effect of talent configuration (TC) properties, ($r_{\rm TA}$, $r_{\rm TD}$) on capital dynamics of the TLS model in BA and WA networks for the short-time scales $t\in[0,100]$.
\begin{figure}[]
  \includegraphics[width=\columnwidth]{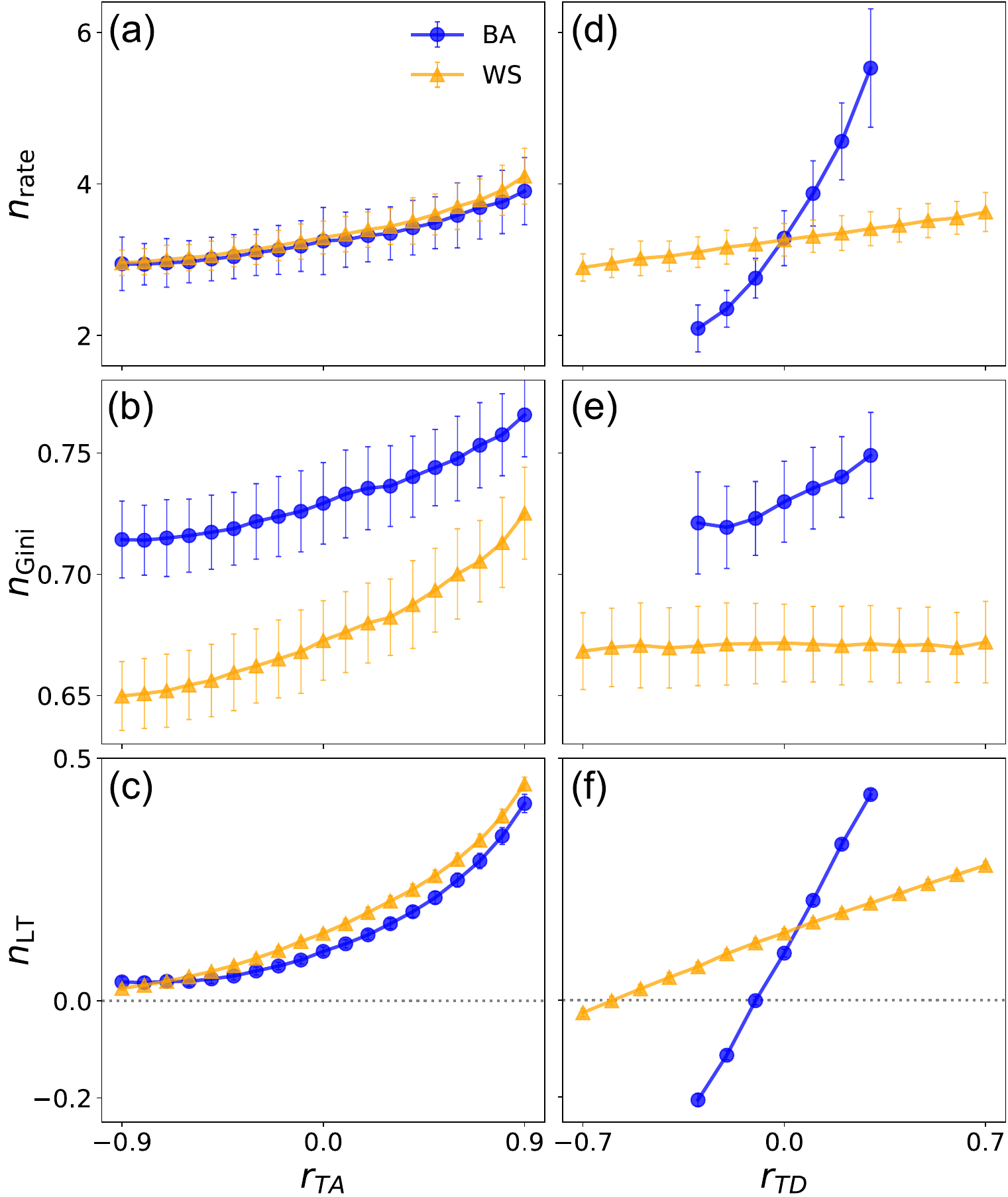}
    \caption{TC effects on three economic indices in BA and WS networks: $n_{\rm rate}$, $n_{\rm Gini}$, and $n_{LT}$ against $r_{\rm TA}$ for (a)-(c), and $r_{\rm TD}$ for (d)-(f). Dots (bars) show the average values (standard deviation) of numerical results at time $t=100$ with $2^{10}$ realizations. The BA network is generated by one additional link attachment per node, and the WS network is generated with the rewiring probability $p_{\rm re}=0.1$. Both networks have the same mean degree $\bar{k}=2$. All simulations are performed for $(N,r,g,b,J)=(10^4,2,0.1,0.1,0.1)$ and $T_i\sim\mathcal{N}(0.6,0.1^2)$.}
    \label{fig5-BA+WS-TA+TD}
\end{figure}
%
\begin{figure*}[]
\includegraphics[width=\textwidth]{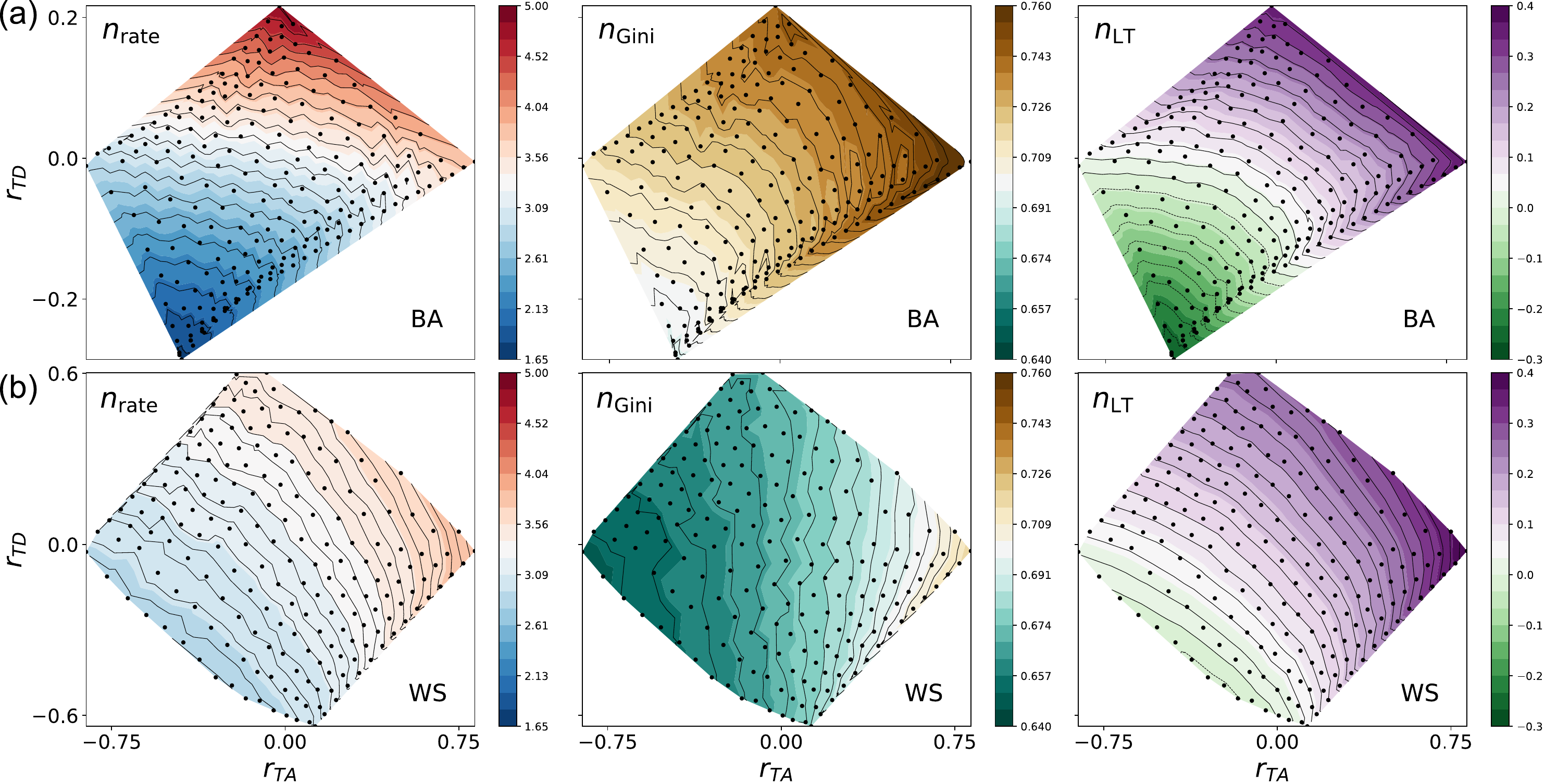}
    \caption{Impact of TC properties on three economic indices. Dots show the ordered pairs of ($r_{\rm TA}$, $r_{\rm TD}$) and colored regions show the expected average values of indices at $t=100$ with $2^7$ realizations. For (a), the BA network is generated by one additional link attachment per node, and for (b), the WS network is generated with the rewiring probability $p_{\rm re}=0.1$. Both networks have the same mean degree $\bar{k}=2$. All simulations are performed for $(N,r,g,b,J)=(10^4,2,0.1,0.1,0.1)$ and $T_i\sim\mathcal{N}(0.6,0.1^2)$.}
    \label{fig6-all_TA+TD}
\end{figure*}

Figure~\ref{fig5-BA+WS-TA+TD} shows TC effects on BA and WS networks as a function of $r_{\rm TA}$ and $r_{\rm TD}$, respectively. The data are obtained in short-term regime and we collect TC for the quasi-pure $r_{\rm TA}$ and $r_{\rm TD}$ cases within the error range of $\epsilon=10^{-2}$. The collected pairs of ($r_{\rm TA}$, $r_{\rm TD}$) are presented in SM, Sec. IV, and Fig.~S5~\cite{SM}. All numerical data for three economic indices of our interests: $n_{\rm rate}$, $n_{\rm Gini}$, and $n_{LT}$ are measured at $t=100$. All time evolution for those in BA and WS networks with the quasi-pure samples of $r_{TA}$ and $r_{TD}$ are also shown in Figs.~S6 and~S7 of SM~\cite{SM}. 

Based on the results in Fig.~\ref{fig5-BA+WS-TA+TD}, we address three remarks: (i) Almost all indices have a positive correlation with $r_{\rm TA}$ and $r_{\rm TD}$, except for (d), $r_{\rm TD}$ vs. $n_{\rm Gini}$ in the WS network. (ii) $n_{LT}$ can not be negative for the quasi-pure $r_{\rm TA}$ case($r_{\rm TD}\sim 0$), while it can for the quasi-pure $r_{\rm TD}$ case($r_{\rm TA}\sim 0$). (iii) For the BA network, $n_{\rm rate}$ explosively increases as $r_{\rm TD}$ increases. In particular, the realization error only depends on $r_{\rm TD}$ for $n_{\rm rate}$. The first remark shows that both $r_{\rm TA}$ and $r_{\rm TD}$ are valid as an criterion for improving three economic indices $n_{\rm rate}$, $n_{\rm Gini}$, and $n_{LT}$. Meanwhile, it can be considered as the trade-off relation between ``growth or meritocracy” and ``equality", which matches the common sense of economy. The second remark shows that the less talent cannot get the more average capital by the $r_{\rm TA}$ effect in the TLS model if there is no correlation between talent and degree for the given network. The third remark shows that $r_{\rm TD}$ effect is critical to the degree heterogeneous network, such as the BA network with the scale-free property.

The ``high degree advantage" combined with high $r_{\rm TD}$ makes the choice and concentration of the highly talented agent, which yields an explosive growth in the system. As a result, $n_{\rm rate}$ of the system strongly depends on performances of a few hub agents and the system volatility increases (even more, higher talent gives higher percentage volatility $\beta$). Therefore, in the TLS model, socioeconomic homophily (high $r_{\rm TA}$) creates a social dilemma between growth and equality, and the hub monopolization by a few highly talented agents (high $r_{\rm TD}$) in a scale-free network makes economic growth highly dependent on their performances. One can also consider the simultaneous effects of both $r_{\rm TA}$ and $r_{\rm TD}$.

As shown in Fig.~\ref{fig6-all_TA+TD}, the contour plots represent both $r_{\rm TA}$ and $r_{\rm TD}$ effects on [Fig.~\ref{fig6-all_TA+TD}(a)] BA and [Fig.~\ref{fig6-all_TA+TD}(b)] WS networks. Most indices have positive correlation between $r_{\rm TA}$ and $r_{\rm TD}$. In the BA network, $n_{\rm rate}$ is more sensitive to $r_{\rm TD}$ than $r_{\rm TA}$ since contour lines almost lie horizontally (along the $r_{\rm TA}$ axis), and $n_{\rm Gini}$ is more sensitive to $r_{\rm TA}$ than $r_{\rm TD}$ since contour lines almost lie vertically (along the $r_{\rm TD}$ axis). In the WS network with small rewiring probability, $p_{\rm re}=0.1$~\cite{small-p_re}, all three indices $n_{\rm rate}$, $n_{\rm Gini}$, and $n_{LT}$ are more sensitive to $r_{\rm TA}$ than $r_{\rm TD}$. Based on these results, one can imply which one, either $r_{\rm TA}$ or $r_{\rm TD}$, is dominant to economic indices and mainly depends on network structure.

\section{\label{summary} Summary and Discussion}

We proposed an agent-based model for capital dynamics with talent, luck, and social interaction (TLS), named the TLS model, where we explored the model by analytical and numerical means. In particular, we showed that the TLS model can be considered as the generalized framework since it covers both the ``talent versus luck" (TvL) model and the Bouchaud-M{\'e}zard (BM) model in the context of the stochastic differential equation form. Inserting talent heterogeneity and interaction in agent-based networks to our model simultaneously, talent configuration (TC) plays a key role in capital dynamics, which was not considered in the BM model. To estimate TC effects systematically, we employed three economic indices: $n_{\rm rate}$ (growth rate), $n_{\rm Gini}$ (inequality), and $n_{LT}$ (meritocratic fairness) to extract TC properties as talent assortativity (TA, $r_{\rm TA}$), and talent-degree correlation (TD, $r_{\rm TD}$).

Our study reveals that TA and TD are positively correlated in three economic indices. In addition, the dominant TC property depends on the network structure. The existing TvL model study suggests that talent requires a supportive environment for full utilization. The environment refers to the total opportunity shared by all agents. We also introduced another environmental factor: the influence of neighboring agents. Unlike the aggregate opportunity, which is a global factor, network interactions are local. An agent's success is significantly influenced by its position within the network and the composition of its neighbors. Network locality restricts the benefits of interaction, and $r_{\rm TA}$ and $r_{\rm TD}$ determine who benefits most. More precisely, high $r_{\rm TA}$ ensures that qualitative benefits (high-talent neighbors) are monopolized by high-talent clusters, and high $r_{\rm TD}$ ensures that quantitative benefits (many neighbors and high-degree advantages) are monopolized by high-talent agents. While growth rates, inequality, and meritocratic fairness increase, high $r_{\rm TA}$ and high $r_{\rm TD}$ can initiate a kind of cartel and centralization effects, respectively. In particular, selective interaction by high-talent clusters under the limited opportunity aggregate condition with even a not so large probability for lucky events can promote the formation of a meritocratic society; see Fig. S8 in SM~\cite{SM}. Our findings provide some insights into socioeconomic homophily (clustering by similar socioeconomic characteristics) and resource concentration among elites. Conversely, socioeconomic integration might appear crucial for reducing inequality.

For future studies, it would be interesting to quantify the long-term impact of TC on the Gini coefficient (beyond the short-term analysis, $t\in[0, 100]$~\cite{long-term}). In addition, the TC effect is generalized onto scale-free networks with various degree decay exponents and explored for the interplay with the shortcut effect in WS networks for different $p_{\rm re}$ and $\bar{k}$ values are promising avenues. The limitations of our study lie in the static nature of interactions. Future extensions could incorporate (1) synergetic interactions increasing the total capital and (2) competitive ones favoring highly talented agents in capital transfer. Alternatively, the consideration of time-dependent talent changes could model agent productivity improvements. For the zero-sum capital scenarios over time, an individual's capital growth only depends on interaction-based transfers. The yard sale (YS) model~\cite{chakraborti2000statistical,hayes2002computing} highlights the role of capital stock in such fixed-sum scenarios, further emphasized by Boghosian {\it et al.}~\cite{boghosian2017oligarchy} for the discontinuous Gini coefficient variation under ``wealth attained advantage". Finally, the most recent paper by Lee and Lee~\cite{lee2023scaling} generalized the YS model on networks and suggested exploring potential phase transitions and scaling behaviors in capital dynamics. Investigating the phase diagram of capital dynamics and condensation transitions could be another fruitful direction.

\begin{acknowledgments} 
This research was supported by the Basic Science Research Program through the National Research Foundation of Korea (NRF) (KR) [Grant No. NRF-2020R1A2C1007703~(J.H., M.H.) and No. NRF-2022R1A2B5B02001752~(J.H., H.J.)]. J.H. thanks Seungwoong Ha for introducing the interesting research topic at the journal club.
\end{acknowledgments} 

\begin{appendix}
\setcounter{figure}{0}
\setcounter{table}{0}
\renewcommand{\thefigure}{A\arabic{figure}}
\renewcommand{\thetable}
{A\arabic{table}}

\section{\label{TvL} Talent versus Luck (TvL) model}
We here briefly review how to study the TvL model. As the discrete time $t$ elapses, the probability that an agent with talent $T$ has a capital level $L=m-n$, obeys the trinomial series as follows: 
\begin{align}
    P_{t}(m,n) = a^mb^n[1-a-b]^{t-m-n}\frac{t!}{m!n!(t-m-n)!},
    \label{eq-P_t}
\end{align}
where $a=gp_g(T)$.
For this case, let us consider four basic statistical quantities of $P_t(L=m-n)$ as a function of time $t$, mean ($\mu_{L}$), standard deviation ($\sigma_{L}$), skewness ($\mathcal{S}_{L}$), and kurtosis ($\mathcal{K}_{L}$):
\begin{flalign}
  \mu_{L}(t) &= t(a-b),\\
  \sigma_{L}(t) &= \sqrt{t}\sqrt{(a+b)-(a-b)^2}, \\
  \mathcal{S}_{L}(t) &= \frac{(a-b)\left[2(a-b)^2-3(a+b)+1\right]}{\sqrt{t}\left[(a+b)-(a-b)^2\right]^{3/2}}, \\
  \mathcal{K}_{L}(t) &= 3+\frac{1}{t}\left[-6+\frac{12ab+(a+b)-(a-b)^2}{[(a+b)-(a-b)^2]^2}\right].
   \label{eq-four-stats}
\end{flalign}
As $t\to\infty$, skewness $\mathcal{S}_{L} \sim t^{-1/2}\to 0$ and kurtosis $\mathcal{K}_{L}\sim 3+ t^{-1}\to 3$, just as the same values for the normal distribution. It is because the sum of $\Delta L=\{-1,0,+1\}$ independently drawn from the same probability mass function follows the central limit theorem.

\section{\label{SDE-TvL} Stochastic Differential Equation (SDE) for the TvL model}

For the TvL model as the trinomial series $P_t(L)$, 
$\mu_{L}\sim t, ~\sigma_{L}\sim t^{1/2},~ \mathcal{S}_{_L}\to 0~, \mbox{and}~\mathcal{K}_{_L}\to 3$
as $t\to\infty$. Thus, the motion of the capital level $L$ can be approximated by the Brownian motion with drift $v_{L}$ and volatility $\theta_{L}$, as $v_{L}\equiv\mu_{L}/t~\mbox{and}
~\theta_{L}\equiv\sigma_{L}/\sqrt{t}$. As a function of $T$, 
\begin{align} 
v_L(T) &= gp_g(T)-b, \\
\theta_L(T) &= \sqrt{[gp_g(T)+b]-[gp_g(T)-b]^2}.
 \label{eq-mu+sigma}
\end{align}
Then we can rewrite the model to the SDE for the capital level $L$ per agent with talent $T$ as follows:
\begin{align}
dL=v_L(T)dt+\theta_L(T)dW_t,
\label{eq-SDE}
\end{align}
where $dt$ is the time interval, $W_t$ is the Wiener process, $v_L$ is drift, and $\theta_L$ is volatility. To represent this SDE for capital $C$, we assume that 
\begin{align}
dC=\alpha Cdt+\beta CdW_t,
\label{eq-SDE4C}
\end{align}
where the percentage drift $\alpha$ and the percentage volatility $\beta$ can be written by the drift $v_L$ and volatility $\theta_L$. In addition, we use the identity of $C$ and $L$ as well as It{\^o} calculus, such that
\begin{align} 
    \beta &= \theta_L \ln{r}, \\ 
    \alpha &= v_L \ln{r}+\frac{1}{2}\beta^2.
    \label{eq-beta+alpha}
\end{align}
Hence, the SDE of the TvL model can be written with four parameters $(r,g,b,T)$ as follows:
\begin{align}
    dC(t)=\alpha(T) C(t)dt+\beta(T) C(t)dW_t,
    \label{eq-SDE_TvL}
\end{align}
which is the well-known geometric Brownian motion (GBM). If the talent distribution follows the normal distribution, $T_i\sim\mathcal{N}(\mu,\sigma^2)$, Eq.~\eqref{eq-SDE_TvL} becomes
\begin{align}
dC_i(t)=\alpha(T_i)C_i(t)dt+\beta(T_i)C_i(t)dW_{t,i},
\label{eq-SDE_T_i}
\end{align}
where $C_i$ is the capital of an agent $i$, $T_i$ is the talent of an agent $i$, and $W_{t,i}$ is the Wiener process of $i$ at time $t$. It is noted that Wiener processes for all agents are independent and follow the It{\^o} interpretation.

To determine this continuous SDE is in a good approximation for the TvL model, the numerical results of the discrete version should be compared with those of the continuous one. It can be said that the SDE is in a good approximation if the $L$ distributions are identical in both versions, see
Supplemental Material (SM), Sec.~I and Fig.~S1~\cite{SM}. 

\section{\label{power-law-TvL} Power-law decay of capital distribution in the TvL model}

We here discuss the capital distribution $\rho$ and its power-law behavior in the TvL model. It is noted that the capital distribution of the GBM is not any power-law but log-normal. Therefore, the capital distribution in the TvL model also follows the Gaussian sum of the log-normal distribution. 

However, the normalized capital $c\equiv C/\bar{C}$ distribution in both the GBM and the TvL model show the power-law behavior. The Fokker-Planck equation of the GBM with talent $T$ can be written as
\begin{align}
    \frac{\partial\rho(c,t)}{\partial t}=\frac{1}{2}\frac{\partial^2}{\partial c^2}[\beta^2c^2\rho(c,t)],
    \label{eq-FP-GBM}
\end{align}
where the equilibrium condition is $\partial \rho(c,t)/\partial t=0$, so that Eq.~\eqref{eq-FP-GBM} becomes the well-known Cauchy-Euler equation. Thus, the equilibrium solution of $c$ distribution for the GBM is 
$$\rho_{eq}(c)=Ac^{-1}+Bc^{-2},$$ 
where we check two constants $A$ and $B$ by numerical means. In the long-term regime, we observe that $\rho(c,\infty)\to\rho_{\rm eq}(c)$ and $B=0$; see SM, Sec. S2, and Fig. S2~\cite{SM}. Therefore, the GBM follows a power-law as $\rho(c)\sim c^{-\gamma}$ with its power-law exponent $\gamma=1$. 

Based on Eq.~\eqref{eq-FP-GBM}, the capital distribution converges to $c^{-1}$, which is no longer as a function of $T$. Therefore, for all agents with the talent distribution $T_i\sim\mathcal{N}(\mu,\sigma^2)$ also follows $\rho_{eq}(c)\sim c^{-1}$ in the limit of $t\to\infty$. Since the Pareto distribution with $1<\gamma<2$ gives Gini coefficient (the index of the inequality) as 1, and both two models always converge to a global condensation state. It implies that a single agent monopolize almost the entire capital of the system as $t\to\infty$.

\section{\label{MF-TLS} Mean-field results of the TLS model}

We here discuss the analytic result of the mean-field TLS model. To do so, we need to revisit the GBM and the mean-field BM model, in the context of the SDE as follows:
\begin{align}
    dC_i &= \alpha C_idt+\beta C_idW_{t,i}, \\
    dC_i &= \alpha C_idt+\beta C_idW_{t,i}-Jdt(C_i-\bar{C}), 
    \label{eq-GBM+MF-BM}
\end{align}
where $\alpha$ and $\beta$ are constants. The mean capital in both models are the same as $\langle{C}\rangle=C_0 e^{\alpha t}$. Summing up all equations in Eq.~\eqref{eq-GBM+MF-BM} over the agent index $i$, the interaction term is canceled out, and the ensemble average of $dC$ just follows the sum of the same GBM. In other words, the mean-field interaction does not change the mean capital, which is also the same as the GBM so far. This property also remains when the interaction term take the general form as: 
\begin{align}
\sum_{j(\neq i)}J_{ij}C_j-\sum_{j(\neq i)}J_{ji}C_i.
   \label{eq-interaction}
\end{align}
Using Eq.~\eqref{eq-GBM+MF-BM}, we construct the SDE of the normalized capital in the mean-field BM model, so that:
\begin{align}
    dc=J(1-c)dt+\beta cdW_{t}.
    \label{eq-SDE_MF-BM}
\end{align}
The corresponding Fokker-Planck equation 
becomes
\begin{align}
    \frac{\partial\rho}{\partial t}=-\frac{\partial}{\partial c}\left[J(1-c)\rho\right]+\frac{1}{2}\frac{\partial^2}{\partial c^2}\left[(\beta c)^2\rho\right].
   \label{eq-FP_MF-BM}    
\end{align}
Then the equilibrium distribution of normalized capital $c$ that satisfies the condition of $\partial \rho/\partial t=0$ is known as
\begin{align}
    \rho_{eq}(c)=\frac{(\frac{2J}{\beta^2})^{1+\frac{2J}{\beta^2}}}{\Gamma(1+\frac{2J}{\beta^2})} \cdot c^{-(2+\frac{2J}{\beta^2})}e^{-\frac{2J}{\beta^2c}},
    \label{eq-rho_eq}
\end{align}
where $\Gamma(x)$ is a Gamma function and $\rho_{eq}(c)\sim c^{-\gamma}$ with the power-law tail exponent of $\gamma=2+2J/\beta^2$. It is noted that this result comes from $\langle C\rangle=C_0e^{\alpha t}$.

Consider the relationship between the TvL model and the mean-field TLS model as follows:
\begin{align}
    dC_i &= \alpha_iC_idt+\beta_iC_idW_{t,i},\\
    dC_i &= \alpha_iC_idt+\beta_iC_idW_{t,i}-Jdt\left(C_i-\bar{C}\right), 
    \label{eq-TvL+MF-TLS}
\end{align}
where $\alpha_i=\alpha(T_i)$ and $\beta_i=\beta(T_i)$. 
For our case, the talent distribution follows the normal distribution, and the mean capital of the TvL model is the Gaussian sum of $e^{\alpha(T)t}$. In the short-term regime, this mean capital does not show exponential growth, see SM, Sec. S2 and Fig.~S3~\cite{SM}. Summing all second equations over agent index $i$, the interaction term is still canceled out. 

Nevertheless, the mean capital in the mean-field TLS model is not equal to that in the TvL model because the sum of $dC_i$ is the sum of different GBMs. Capital transfer by the mean-field interaction $-Jdt(C_i-\bar{C})$ makes relative changes in agent capitals, but it does not change the total capital at that time. However, these relative capital changes between talent-heterogeneous agents makes difference in capital growth eventually. This talent heterogeneity makes a system more complex. 
 
In order to solve Eq.~\eqref{eq-TvL+MF-TLS}, $\langle C\rangle=C_0e^{\tilde{\alpha}t}$ is assumed, where $\tilde{\alpha}$ is constant. Then the SDE for the normalized capital $c_i$ with talent $T_i$ becomes
\begin{align}
    dc_i=(J-K_ic_i)dt+\beta_ic_idW_{t,i},
    \label{eq-SDE_MF-TLS}
\end{align}
where $K_i=J+\tilde{\alpha}-\alpha_i$. 

The Fokker-Planck equation of Eq.~\eqref{eq-SDE_MF-TLS} becomes
\begin{align}
    \frac{\partial\rho_i}{\partial t}=-\frac{\partial}{\partial c_i}\left[\{J-K_ic_i\}\rho_i\right]+\frac{1}{2}\frac{\partial^2}{\partial c_i^2}\left[(\beta_i c_i)^2\rho_i\right],
    \label{eq-FP_MF-TLS}
\end{align}
where $\partial \rho_i/\partial t=0$ is the equilibrium condition. The equilibrium solution of the normalized capital distribution $\rho_{eq,i}$ for the $T_i$ talented group is as follows:
\begin{align}
    \rho_{eq,i}(c_i)=\frac{(\frac{2J}{\beta_i^2})^{1+\frac{2K_i}{\beta_i^2}}}{\Gamma(1+\frac{2K_i}{\beta_i^2})}\cdot c_i^{-(2+\frac{2K_i}{\beta_i^2})}e^{-\frac{2J}{\beta_i^2c_i}}.
    \label{eq-rho_eq-i}
\end{align}

\begin{figure}[]
\includegraphics[width=0.85\columnwidth]{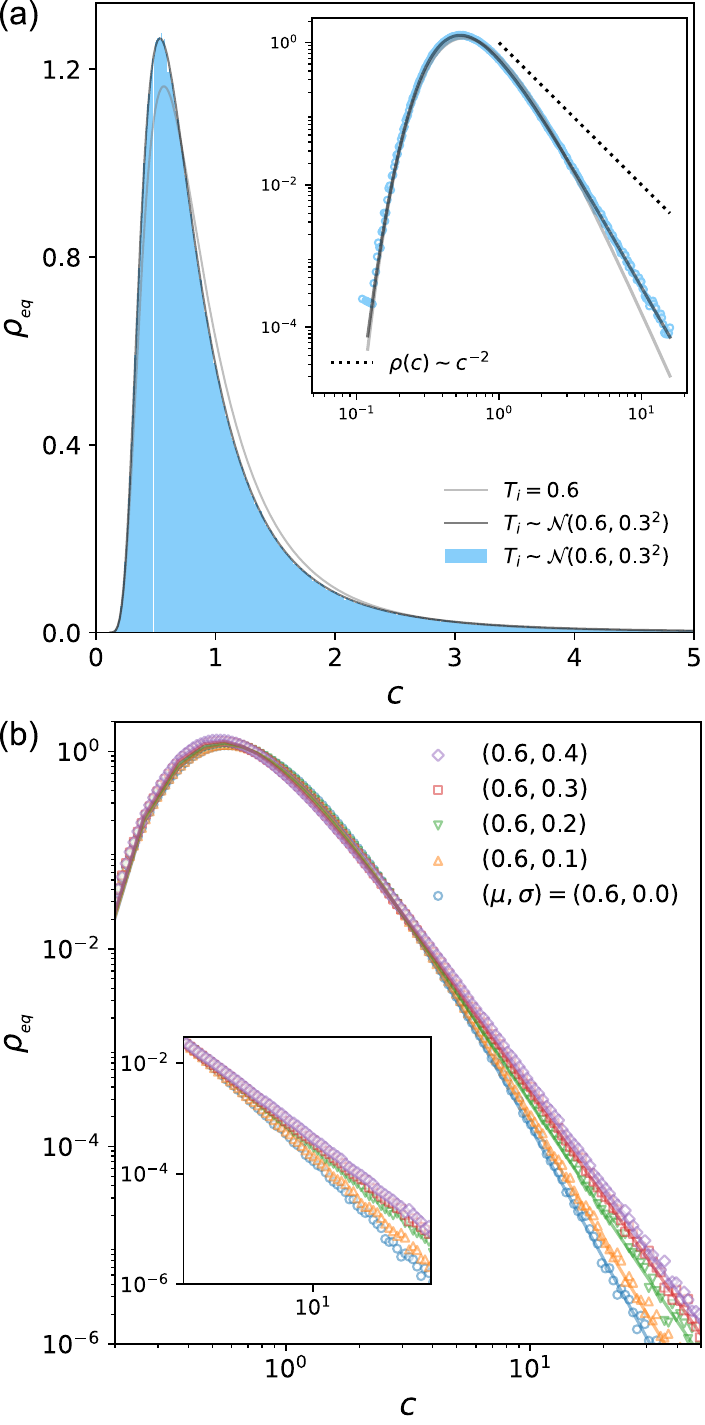}
    \caption{Equilibrium normalized capital distribution in mean-field TLS model. (a) The colored area shows the result of the normalized capital histogram as $t\to\infty$. Gray and black solid lines are the analytic solutions of the mean-field BM model~\eqref{eq-SDE_MF-BM} and the mean-field TLS model~\eqref{eq-SDE_MF-TLS} for $T_i=0.6$ and $T_i\sim\mathcal{N}(0.6,0.3^2)$, respectively. The inset shows the double-logarithmic scaled plots of the main plots. (b) For a variety of the standard deviations $\sigma$ values with the same $\mu$, the decaying behavior of $\rho_{eq}$ is double logarithmically plotted against $c$, and the diffrent portion near $c=10^1$ is shown in the inset. All simulations are performed for $(N,r,g,b,J)=(2\times 10^6,2,0.1,0.1,0.1)$.}
    \label{figA1-rho_eq}
\end{figure}

\renewcommand{\arraystretch}{1.2}
\begin{table*}[t]
\caption{Main results of wealth dynamics models. Here $\alpha_i=\alpha(T_i)$, $\beta=\beta(T_i)$, and the talent $T$ satisfies the normal distribution as $p(T)=\frac{1}{\sigma\sqrt{2\pi}}e^{-\frac{(T-\mu)^2}{2\sigma^2}}$. An adjusting factor $\delta$ satisfies $0\leq\delta\leq\langle{\frac{2J}{\beta^2}}\rangle$, depending on $p(T)$.}
\begin{center}
\begin{tabular}{ c c c c } 
\hline\hline
Model & Growth & $\rho_{eq}(c)$ & $\gamma$ \\
\hline
\multirow{2}{14em}{\centering{GBM}:\\ {$dC_i=\alpha C_idt+\beta C_idW_{t,i}$}} & 
\multirow{2}{14em}{\centering{Exponential}:\\ $\langle{C}\rangle=C_0e^{\alpha t}$} & \multirow{2}{16em}{\centering{$\sim c^{-1}$}} & \multirow{2}{*}{$1$} \\
& & &\\
\hline
\multirow{2}{14em}{\centering{TvL model}:\\ {$dC_i=\alpha_i C_idt+\beta_i C_idW_{t,i}$}} & \multirow{2}{14em}{\centering{Non-exponential}:\\ $\langle{C}\rangle=\int_{-\infty}^{\infty}p(T)C_0e^{\alpha(T)t}dT$} & \multirow{2}{16em}{\centering{$\sim c^{-1}$}} & \multirow{2}{*}{$1$} \\
& & & \\
\hline
\multirow{3}{14em}{\centering{Mean-field BM model}:\\ {$dC_i=\alpha C_idt+ \beta C_idW_{t,i}$} \\{\hspace{1cm}$-Jdt(C_i-\bar{C})$}} & \multirow{3}{14em}{\centering{Exponential}:\\ $\langle{C}\rangle=C_0e^{\alpha t}$} & \multirow{3}{16em}{\centering{$f(c;1+\frac{2J}{\beta^2},\frac{2J}{\beta^2})$}$^{a}$} & \multirow{3}{*}{$2+\frac{2J}{\beta^2}$} \\
& & \\
& & \\
\hline
\multirow{3}{14em}{\centering{Mean-field TLS model}:\\ {$dC_i=\alpha_i C_idt+ \beta_i C_idW_{t,i}$}\\ {\hspace{1cm}$-Jdt(C_i-\bar{C})$}} & \multirow{3}{14em}{\centering{Exponential}:\\ $\langle{C}\rangle=C_0e^{\tilde{\alpha}t}$,\\ $[\alpha(0)<\tilde{\alpha}<\alpha(1)$]} & \multirow{3}{16em}{\centering{$\int_{-\infty}^{\infty}p(T)f(c;1+\frac{2K(T)}{\beta(T)^2},\frac{2J}{\beta(T)^2})dT$}} & \multirow{3}{*}{$2+\langle{\frac{2J}{\beta^2}}\rangle-\delta$} \\
& & & \\
& & & \\
\hline\hline
\end{tabular}
\end{center}
\begin{flushleft}
$^a${$f(c;A,B)\equiv\frac{B^{A}}{\Gamma(A)}c^{-(1+A)}e^{-B/c}$ denotes the inverse-Gamma distribution.}
\end{flushleft}
\label{table:1}
\end{table*}

If $T_i\sim\mathcal{N}(\mu,\sigma^2)$ is considered, Eq.~\eqref{eq-rho_eq-i} can be
integrated over all agents' $T_i$:
\begin{align}
    \rho_{eq}(c)=\int_{-\infty}^{\infty}\frac{e^{-\frac{(T-\mu)^2}{2\sigma^2}}}{\sigma\sqrt{2\pi}}\cdot\frac{(\frac{2J}{\beta^2})^{1+\frac{2K}{\beta^2}}}{\Gamma(1+\frac{2K}{\beta^2})}\cdot c^{-(2+\frac{2K}{\beta^2})}e^{-\frac{2J}{\beta^2c}}dT.
    \label{eq-rho_eq-all}
\end{align}
By the numerical simulation of the SDE for the mean-field TLS model, we check the assumption and the solution. To do so, we define $\alpha_{\rm ex}$ as
\begin{align}
    \alpha_{\rm ex}=\frac{\ln{(\langle{C}\rangle/C_0)}}{t}.
    \label{eq-alpha_ex}
\end{align}
If the mean capital of the system grows exponentially, $\alpha_{\rm ex}$ must be a constant $\tilde{\alpha}$. The time evolution of $\alpha_{\rm ex}$ is tested in SM; see Fig. S4~\cite{SM}. The exponent $\alpha_{\rm ex}$ is deeply related with the correlation between $\alpha$ and $c$ because the sum of Eq.~\eqref{eq-SDE_MF-TLS} ensures that $\tilde{\alpha}=\langle{\alpha c}\rangle$, provided that the exponential growth is true (even for general interaction matrix $J_{ij}$ cases) as drawn in Table~\ref{table:1} and Fig.~\ref{figA1-rho_eq}.

As summarized in Table~\ref{table:1}, for the power-law behaviors on $\rho_{eq}(c)$ in mean-field models, the power-law tail exponent $\gamma$ in the mean-field BM model only depend on $\beta(\mu)^2$ for a given parameter set $(r,g,b,J)$. This is because there is no talent difference between agents. Thus, 
\begin{align}
\gamma_{\rm BM}=2+2J/\beta(\mu)^2,
\label{eq-gamma_BM}
\end{align}
where $\sigma=0$ and $T$ is a constant. 
However, in the mean-field TLS model, it is complicated because $\sigma\neq0$, so that talent $T$ is not a constant but a variable of the normal distribution. Under such a condition, not only average talent $\mu$ but also talent heterogeneity $\sigma$ can influence the power-law tail exponent $\gamma$. We empirically find that 
\begin{align}
\gamma_{\rm TLS}=2+\langle{2J/\beta^2}\rangle-\delta(\mu,\sigma),
\label{eq-gamma_TLS}
\end{align}
where $\langle{\cdots}\rangle$ means the ensemble average and $\delta$ is a adjusting factor function of $(\mu,\sigma)$. How the talent heterogeneity $\sigma$ influence the power-law tail exponent $\gamma$ is shown in Fig.~\ref{figA1-rho_eq}.

\end{appendix}

\bibliographystyle{apsrev4-2}
\bibliography{ref-PRE-TLS}

\begin{thebibliography}{23}
\bibitem [{sma()}]{small-p_re}%
  \BibitemOpen
  \href@noop {} {}\bibinfo {note} {For sufficiently small rewiring probability $p_{\rm re}=0.1$ makes the WS network have more lattice-like property like cycle network.}\BibitemShut {Stop}%
\bibitem [{lon()}]{long-term}%
  \BibitemOpen
  \href@noop {} {}\bibinfo {note} {Some preliminary results have been tested as our future study. As shown in SM, Fig. S8~(c)~\cite{SM}, $n_{\rm Gini}$ converges to 1 after long time later, regardless of $r_{TA}$ value when the network is very sparse as $\bar{k}=2$. It seems to TA effects only play a role of delaying global wealth condensation. However, one may consider that the final destination of system inequality can differ for $r_{\rm TA}$ values in sufficiently dense networks.}\BibitemShut {Stop}%
\end{thebibliography}

\clearpage
\begin{widetext}
\section*{\bf Supplemental Material for ``Interplay of network structure and talent configuration on wealth dynamics''}
\maketitle

\setcounter{section}{0}
\setcounter{equation}{0}
\setcounter{figure}{0}
\setcounter{table}{0}
\setcounter{page}{1}
\makeatletter

\renewcommand{\thesection}{S\arabic{section}}
\renewcommand{\theequation}{S\arabic{equation}}
\renewcommand{\thefigure}{S\arabic{figure}}
\renewcommand{\thetable}{S\arabic{table}}
\renewcommand{\bibnumfmt}[1]{[#1]}
\renewcommand{\citenumfont}[1]{#1}

\newcolumntype{L}[1]{>{\raggedright\arraybackslash}p{#1}}
 
\newcolumntype{C}[1]{>{\centering\arraybackslash}p{#1}}
 
\newcolumntype{R}[1]{>{\raggedleft\arraybackslash}p{#1}}

\section{\label{DvC} Correspondence Check between Discrete and Continuous Models}


\begin{figure*}[b]
    \includegraphics[width=\textwidth]{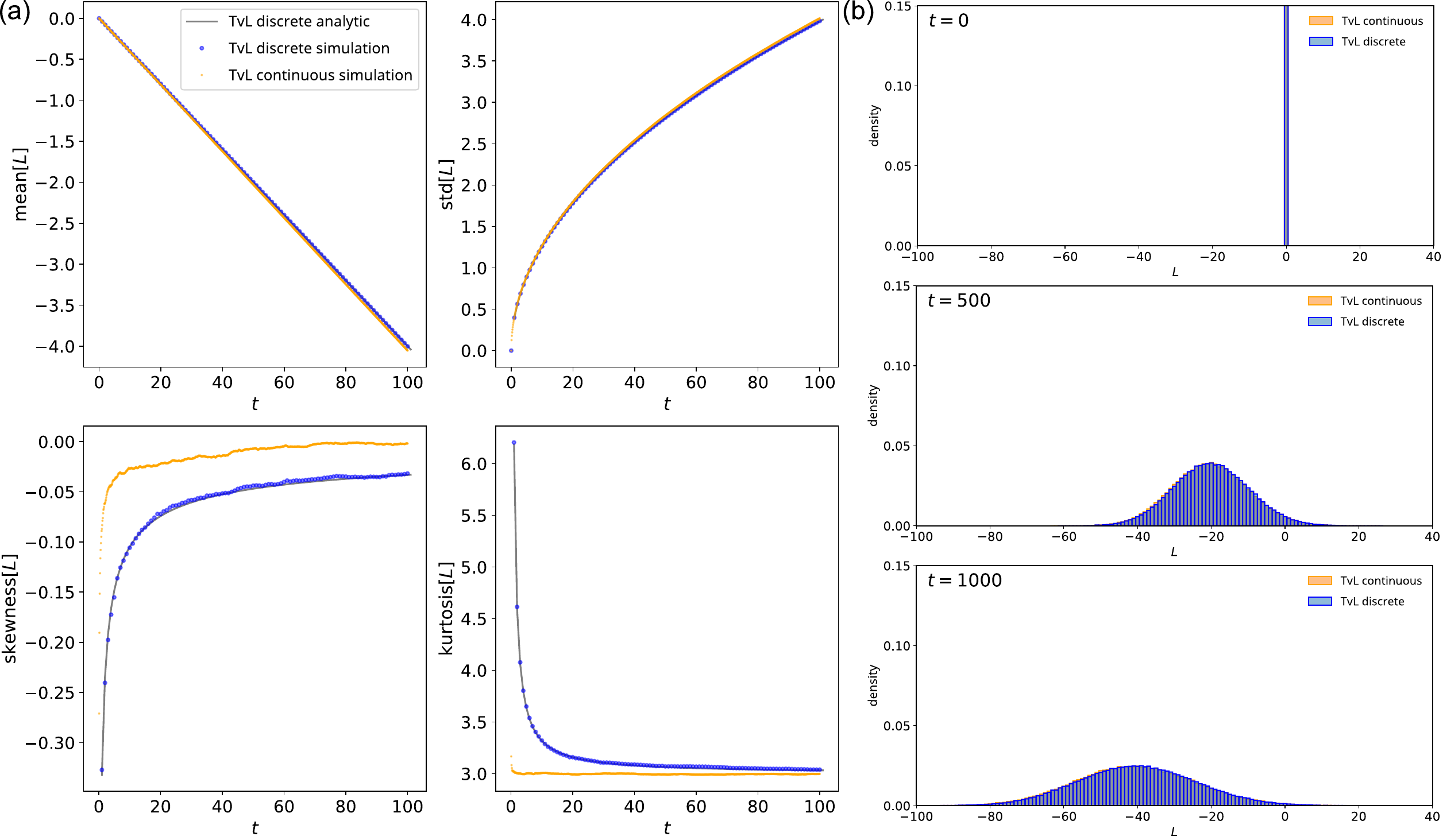}
    \caption{Correspondence check of the TvL model between discrete and continuous versions. The discrete version is numerically simulated by capital change rules, see Fig.~1 in the main text, and the continuous version is numerically obtained by Eq.~(13) in the main text. (a) Four statistics and (b) histograms are shown in discrete and continuous versions of the TvL model. The time interval $\Delta t=1$ is used for discrete version and the time increment $dt=0.1$ is used for continuous version. All simulations are performed for $(N,r,g,b,J)=(10^6,2,0.1,0.1,0.1)$ and $T_i=0.6$.}
    \label{fig-S1}
\end{figure*}

This is the justification of capital dynamics in the talent versus luck (TvL) model between discrete and a continuous versions. In earlier simulation results by Pluchino, {\it et al.}~\cite{pluchino2018talent}, the rate of highly talented agents (talent is higher than $\mu+\sigma$) improved from its first condition ($C_0$) after 80 time step was about to 32~\%. It is also written as $P_T=P(C_i>C_0|T_i>\mu+\sigma)$. To mimic these results, in accordance with our numerical simulation, we employ the simple environmental parameter set as $(r,g,b)=(2,0.1,0.1)$. Under this condition, discrete and continuous versions give $P_{T,dis}=0.253$ and $P_{T,con}=0.294$, respectively, similar as $\sim 32~\%$ by using $(N,\mu,\sigma)=(10^6,0.6,0.1)$.

Since capital $C$ and capital level $L$ satisfy the identity $C=C_0r^L\longleftrightarrow L=\log_{r}(C/C_0)$, we can say that the continuous TvL equation is in good approximation for the discrete TvL model under the condition that both discrete and continuous simulation results give the same capital level $L$ distributions. The $L$ distribution in the TvL model is the Gaussian sum of $L$ distributions of $T$ talented agent groups. If the $L$ distribution in discrete and continuous versions are the same for $T$ talented group, the whole distribution of Gaussian talented group would be the same. Figure~\ref{fig-S1}~(a) shows the four statistics of discrete TvL and continuous TvL simulation results. The statistics of discrete model follows the theoretical prediction (gray line) as we mentioned in main text, see Eq.~(3)--(6). The mean and standard deviation of two are slightly different since $dt$ is finite but almost same. The skewness and the kurtosis of the discrete model more slowly converges but eventually converges to $0$ and $3$ as $t\to\infty$, just as the normal distribution. The $L$ distributions of two are almost the same as shown in Fig.~\ref{fig-S1}~(b). According to this result, we can  consider only the continuous TvL model from now on for the remainders.

\section{Additional Information}

\subsection{Power-law behaviors in GBM and TvL model}
We mention that the equilibrium normalized capital distribution in the TvL model is $\rho_{eq}(c)=Ac^{-1}+Bc^{-2}$, where $A$ and $B$ are constants, which is as a solution of the Fokker-Planck equation. Fig.~\ref{fig-S2} shows that simulation results guarantees $B=0$: for both cases of (a) GBM and (b) the TvL model where $T_i=0.6$ for (a) and $T_i\sim\mathcal{N}(0.6,0.1^2)$ for (b). These two distribution are the same as $\rho_{eq}(c)\sim c^{-\gamma}$ with $\gamma=1$ even though talent distribution are different. This means that talent heterogeneity $\sigma$ does not change $\rho_{eq}(c)$ and $\gamma$ in the TvL model.

\subsection{\label{Non-exp}Non-exponential growth of TvL model}

The growth rate $n_{\rm rate}$ ($\langle{C}\rangle/C_0$) in the TvL model as a function of $(r,g,b,\mu,\sigma)$ is given as follows:
\begin{align}
    \frac{\langle{C}\rangle}{C_0} =& \frac{1}{2\sqrt{2\sigma^2X t+1}}e^{\frac{-\mu^2X t+\mu Y t+\sigma^2Y^2t^2/2}{2\sigma^2X t+1}-Z t}\left[\mathrm{erf}\left(\frac{\mu+\sigma^2Y t}{\sigma\sqrt{4\sigma^2X t+2}}\right)-\mathrm{erf}\left(\frac{\mu+\sigma^2(Y-2X)t-1}{\sigma\sqrt{4\sigma^2X t+2}}\right)\right] \nonumber\\ 
    &+\frac{1}{2}e^{(-X+Y-Z)t}\left[1+\mathrm{erf}\left(\frac{\mu-1}{\sigma\sqrt{2}}\right)\right]+\frac{1}{2}e^{-Zt}\left[1-{\rm erf\left(\frac{\mu}{\sigma\sqrt{2}}\right)}\right],
    \label{TvL-growth_rate}
\end{align}
where
\begin{align*}    
    X=\frac{g^2(\ln{r})^2}{2},\  
    Y=g\ln{r}\left(1+\frac{\ln{r}}{2}+b\ln{r}\right),\ 
    \text{and}\ 
    Z=b\ln{r}\left(1-\frac{\ln{r}}{2}+\frac{b\ln{r}}{2}\right)
\end{align*}

\begin{figure*}[t]
    \includegraphics[width=0.9\textwidth]{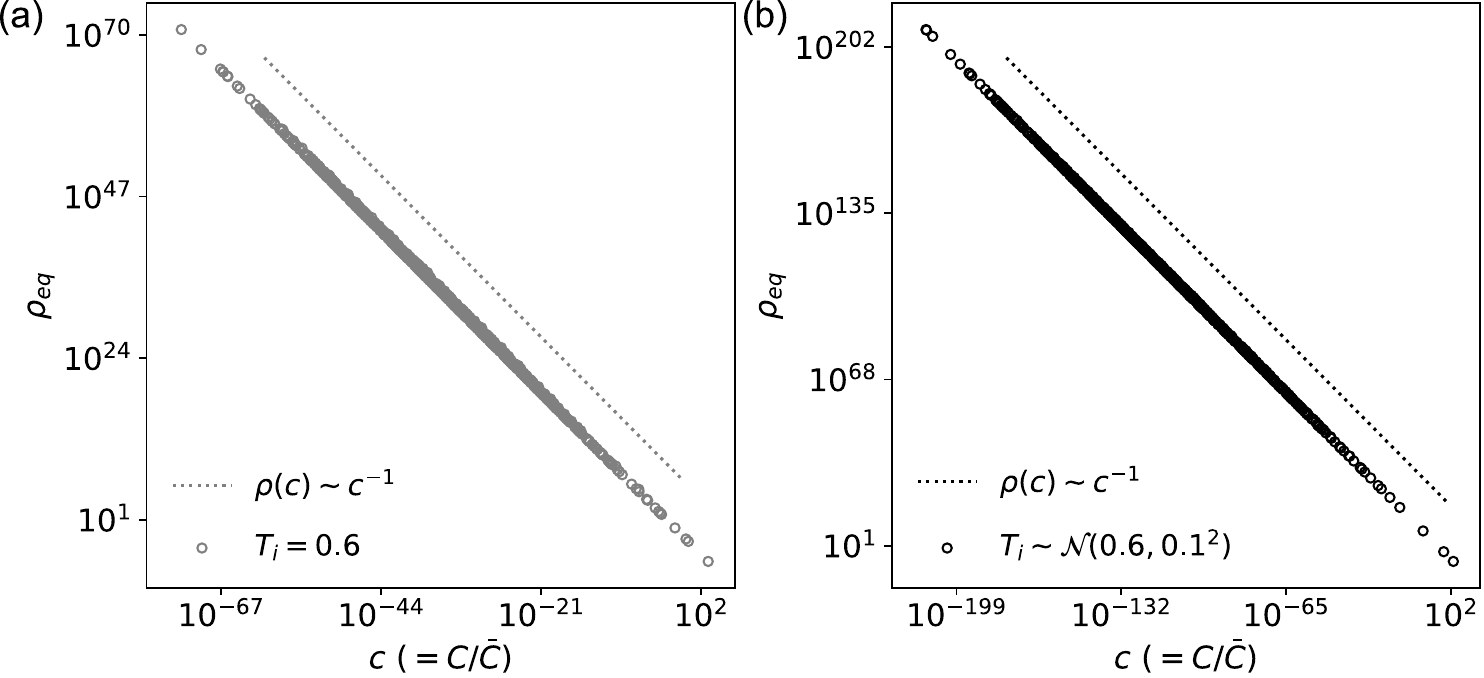}
    \caption{Double-logarithmically scaled plots for equilibrium normalized capital distribution: (a) GBM and (b) the TvL model, where $T_i=0.6$ for (a) and $T_i\sim\mathcal{N}(0.6,0.1^2)$ for (b). All simulations are performed for $(r,g,b)=(2,0.1,0.1)$.}
    \label{fig-S2}
\end{figure*}

\begin{figure*}[t]
    \includegraphics[width=0.95\textwidth]{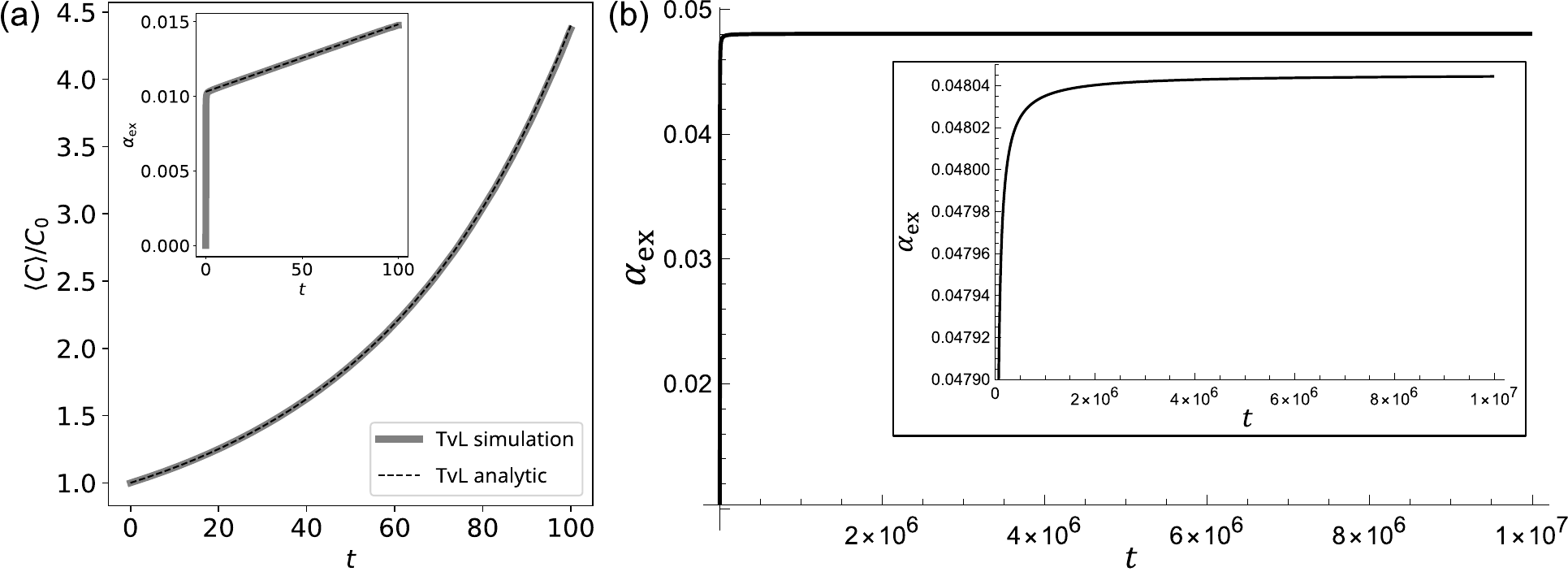}
    \caption{Non-exponential growth in the TvL model. (a) The time evolution of $n_{\rm rate}$ in the TvL model with $N=2^10\times 10^6$. The gray solid line is numerical result and black dashed line is analytic solution represented by Eq.~\eqref{TvL-growth_rate} (b) The time evolution $\alpha_{\rm ex}$ in the TvL model. The TvL motion super-exponentially starts, but eventually converges to the exponential growth with $\alpha_{\rm ex}\simeq \alpha(1)=0.04805$. These results (a) and (b) share the same parameter set $(r,g,b)=(2,0.1,0.1)$ and $T_i\sim\mathcal{N}(0.6,0.1^2)$.}
    \label{fig-S3}
\end{figure*}

As shown in Fig.~\ref{fig-S3} (a), the simulation result fits the analytic solution very well. The inset shows $\alpha_{\rm ex}=\ln{(\langle{C}\rangle/C_0)}/t$ linearly increases as time $t$ elapses. It means the super-exponential growth. However, as $t\to \infty$, it converges to the some value. This converging value is related to the maximum value of $\alpha(r,g,b,T)=\ln{r}(gp_g(T)-b)+\frac{(\ln{r})^2}{2}[(gp_g(T)+b)-(gp_g(T)-b)^2]$. Under the simulation condition of $(r,g,b)=(2,0.1,0.1)$, $T=1$ gives the maximum and its value is $\alpha(2,0.1,0.1,1)=0.04805$, see Fig.~\ref{fig-S3}~(b) as well.

\begin{figure}[h]
    \includegraphics[width=\textwidth]{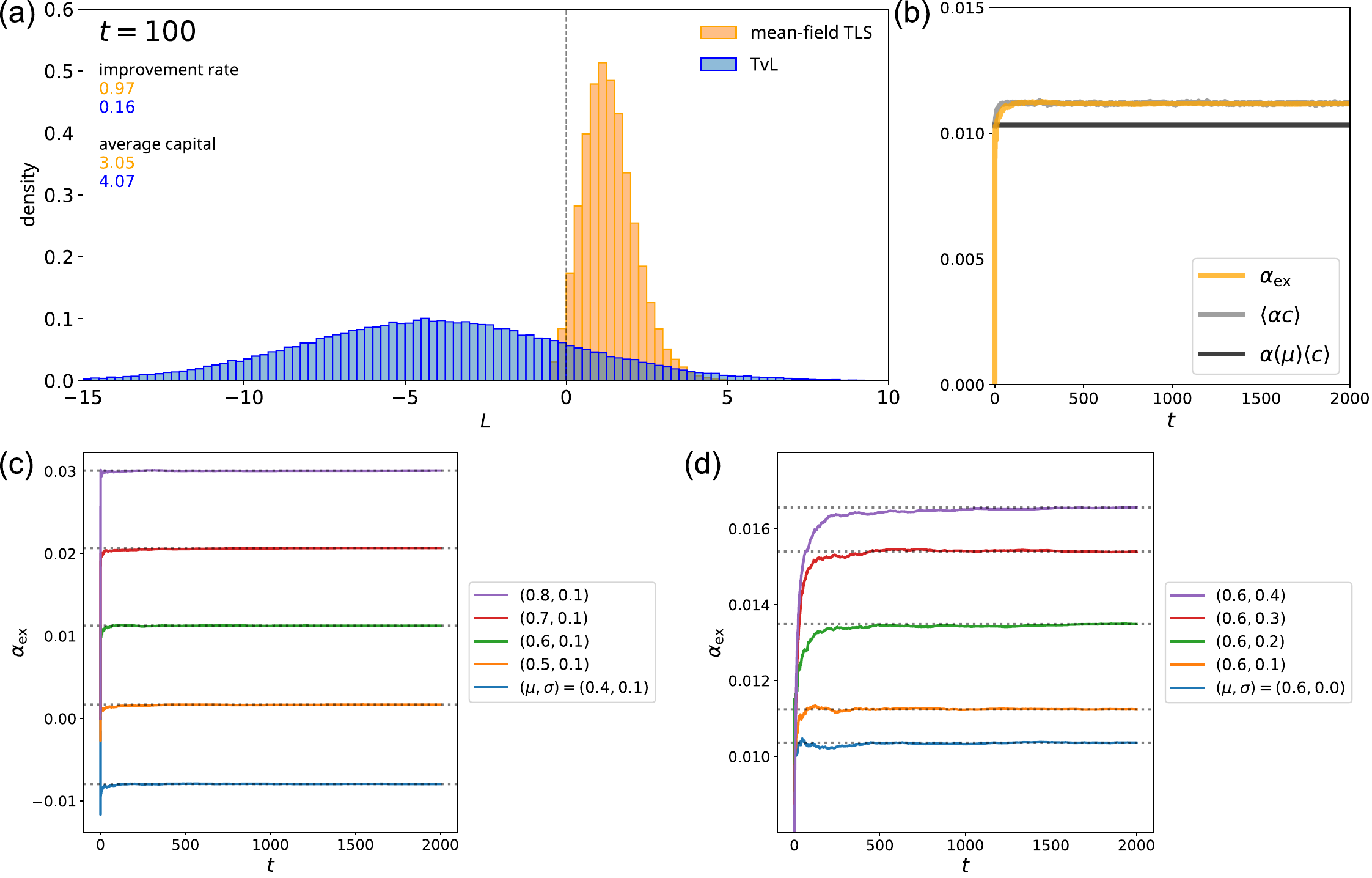}
    \caption{Additional information for mean-field TLS model. (a) The $L$ distribution at $t=100$ with parameter set $(r,g,b,J)=(2,0.1,0.1,0.1)$ and $T_i\sim\mathcal{N}(0.6,0.1^2)$. While the improvement rate in the mean-field TLS model reaches about 1 at $t=100$, in the TvL model, it is not. It means that when the Pareto improvement~\cite{pareto2014manual} is possible in the mean-field TLS model with enough time, whereas in the TvL, it is not. (b) The time evolution of $\alpha_{\rm ex}$ for the mean-field TLS model shown in (a). $\alpha_{\rm ex}$ is related to the correlation between $\alpha$ and $c$. (c) and (d) show how to converge $\alpha_{\rm ex}$ for a variety of parameters: (c) with the fixed standard deviation ($\sigma=0.1$) and (d) with the fixed mean ($\mu=0.6$).}
    \label{fig-S4}
\end{figure}

\subsection{Mean-field TLS model}

We here provide the additional information for the mean-field TLS model. For $(r,g,b,T)=(2,0.1,0.1,0.6)$, $\alpha(2,0.1,0.1,0.6)=0.0103$ and $\beta(2,0.1,0.1,0.6)=0.276$. It roughly implies that the capital change of average talented agent mainly depends on luck than talent 27 times more. The total capital grows exponentially because the mean capital of $T$ talented group is $C_0e^{\alpha(T)t}$, but microscopically, the rate of agents that loses capital than initial capital $C_0$ increases over time.

Fig.~\ref{fig-S4}~(a) shows the $L$ distribution for various models at $t=100$. If $L=\log_r(C/C_0)$ is larger than 0, it means the capital is larger than the initial capital $C_0$. The improvement rate $P_I=P(C>C_0)$ in the TvL model and the mean-field TLS model at $t=100$ are $0.16$ and $0.97$, respectively. The improvement rate in the TvL model decreases over time, whereas in the mean-field TLS model, it increases over time and finally converges to 1. It implies that the Pareto improvement is impossible for the TvL model, whereas it is possible for the mean-field TLS model. That is why we need to employ the interactions to the TvL model, and the interactive model is more persuasive. It is because the system that most people get poorer over time is untenable.

In order to solve the equilibrium normalized capital distribution in the mean-field TLS model, we assume that $\langle{C}\rangle=C_0e^{\tilde{\alpha}t}$. Fig.~\ref{fig-S4}~(b) shows the fact that $\alpha_{\rm ex}$ converges to some value for all cases. This value is related to the correlation between $\alpha$ and $c$ and slightly higher than $\alpha(\mu)\langle{c}\rangle\simeq\langle{\alpha}\rangle\langle{c}\rangle$. It means that the higher $\alpha$ guarantees the higher $c$ since their covariance is not zero even though it is small. The lower and upper bounds of $\tilde{\alpha}$ are $\alpha(r,g,b,T=0)$ and $\alpha(r,g,b,T=1)$, which correspond to the situations that all agents have the minimum effective talent and the maximum one, respectively. 
As shown in Fig.~\ref{fig-S4}~(c) and (d), all $\tilde{\alpha}$ in mean-field models for a variety of talent distributions bounded with the following region: $\alpha(2,0.1,0.1,0)=-0.0477<\tilde{\alpha}<\alpha(2,0.1,0.1,1)=0.04805$. While the power-law exponent does not vary with the talent heterogeneity $\sigma$ in the TvL model, see Fig.~\ref{fig-S2}, it varies with the $\sigma$ in the mean-field TLS model, see Fig.~3~(b) in the main text. As $\sigma$ increases, the power-law tail exponent $\gamma$ decreases. Consequently, in general, relatively richer agents occurs more frequent and the system inequality increases.

\begin{table}[]
\caption{Pseudo-codes for random pair swapping algorithms. In the left panel, a single value control ({\bf Pseudo-code 1}, top) and dual value control ({\bf Pseudo-code 2}, bottom) are shown, and in the right panel, four schematic illustrations are the processes of {\bf Pseudo-code 1}.}
    \includegraphics[width=0.95\textwidth]{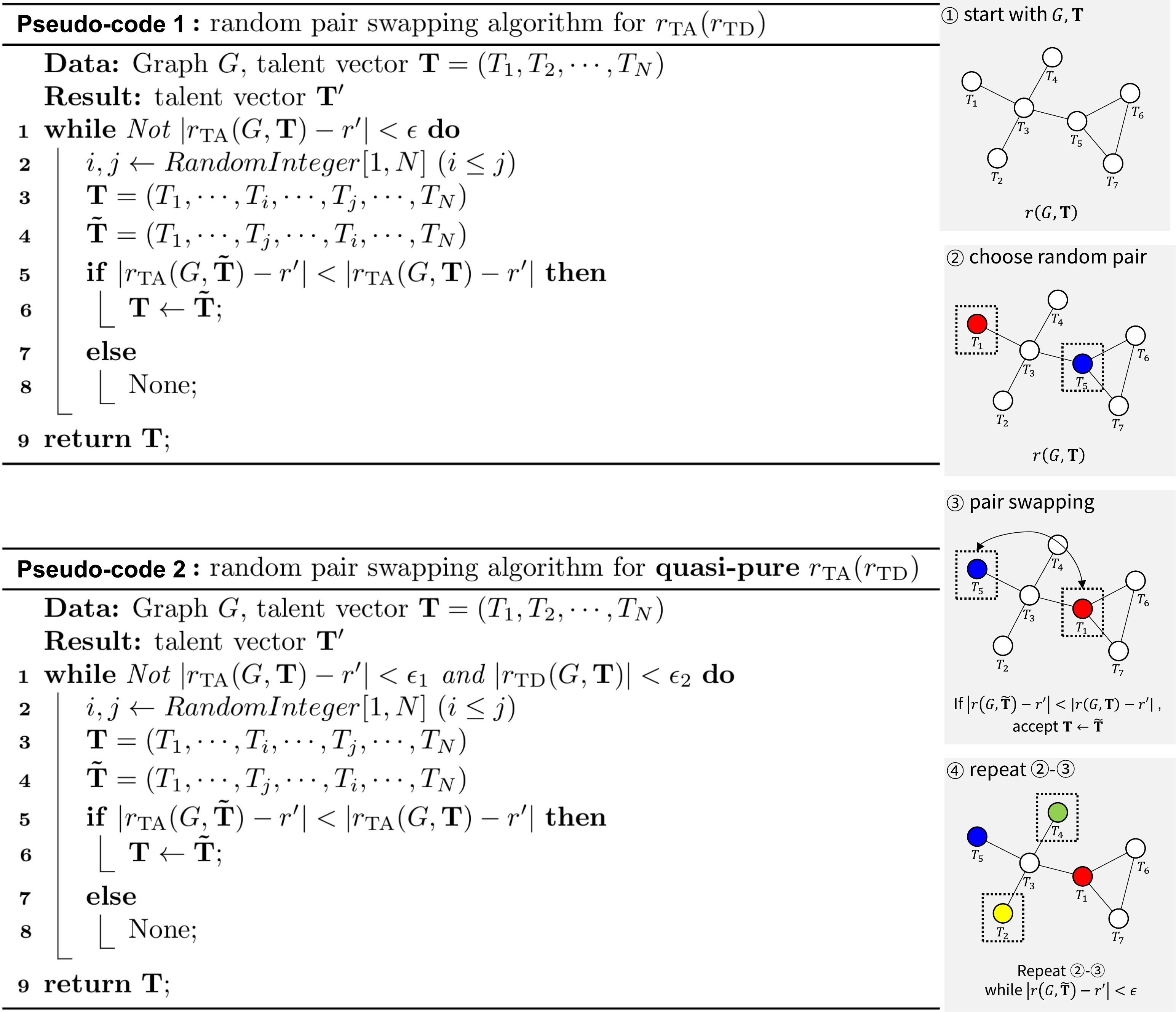}
    \label{table-S1}
\end{table}
\section{\label{RPS}Random pair swapping algorithms}

In order to control both $r_{\rm TA}$ and $r_{\rm TD}$ independently, we employ two algorithms as  shown in Table~\ref{table-S1} with schematic illustrations. Swap of node metadata to minimize (or maximize) the assortativity (similar with {\bf Pseudo-code 1}) suggested by Cinelli {\it et al.}~\cite{cinelli2020network}. In our case, to estimate $r_{\rm TA}$ effect for a given network, we control $r_{\rm TD}$, and vice versa. However, $r_{\rm TA}$ and $r_{\rm TD}$ are correlated with each other under the single pair swapping. So we employ {\bf Pseudo-code 2} in Table~\ref{table-S1} to control the opposite talent configuration property. If $r_{\rm TA}=0$, all agents are uncorrelated with regard to talent assortativity (TA), and vice versa. Therefore, we take the opposite value close to 0 as much as possible to minimize the opposite effect. Thus, we define such a zone as the `quasi-pure zone' by taking $\epsilon=10^{-2}$, so that these samples are `quasi-pure' $r_{\rm TA}$ and $r_{\rm TD}$, respectively.

\section{\label{BA-WS} More information for talent configuration effect}

\subsection{Quasi-pure samples of ($r_{\rm TA}$, $r_{\rm TD}$)}

For Barab{\'a}si-Albert (BA) and Watts-Strogatz (WS) networks~\cite{barabasi1999emergence,watts1998collective} with quasi-pure samples of $r_{\rm TA}$ and $r_{\rm TD}$ as shown in Fig.~\ref{fig-S5}. Quasi-pure samples for BA and WS networks by using {\bf Pseudo-code 2} in Table~\ref{table-S1} with $N\times 100$ times of random pair swapping trials. Talent assortativity (TA, $r_{\rm TA}$) and talent-degree correlation (TD, $r_{\rm TD}$) are correlated under single pair swapping.

\begin{figure*}[]
    \includegraphics[width=0.75\textwidth]{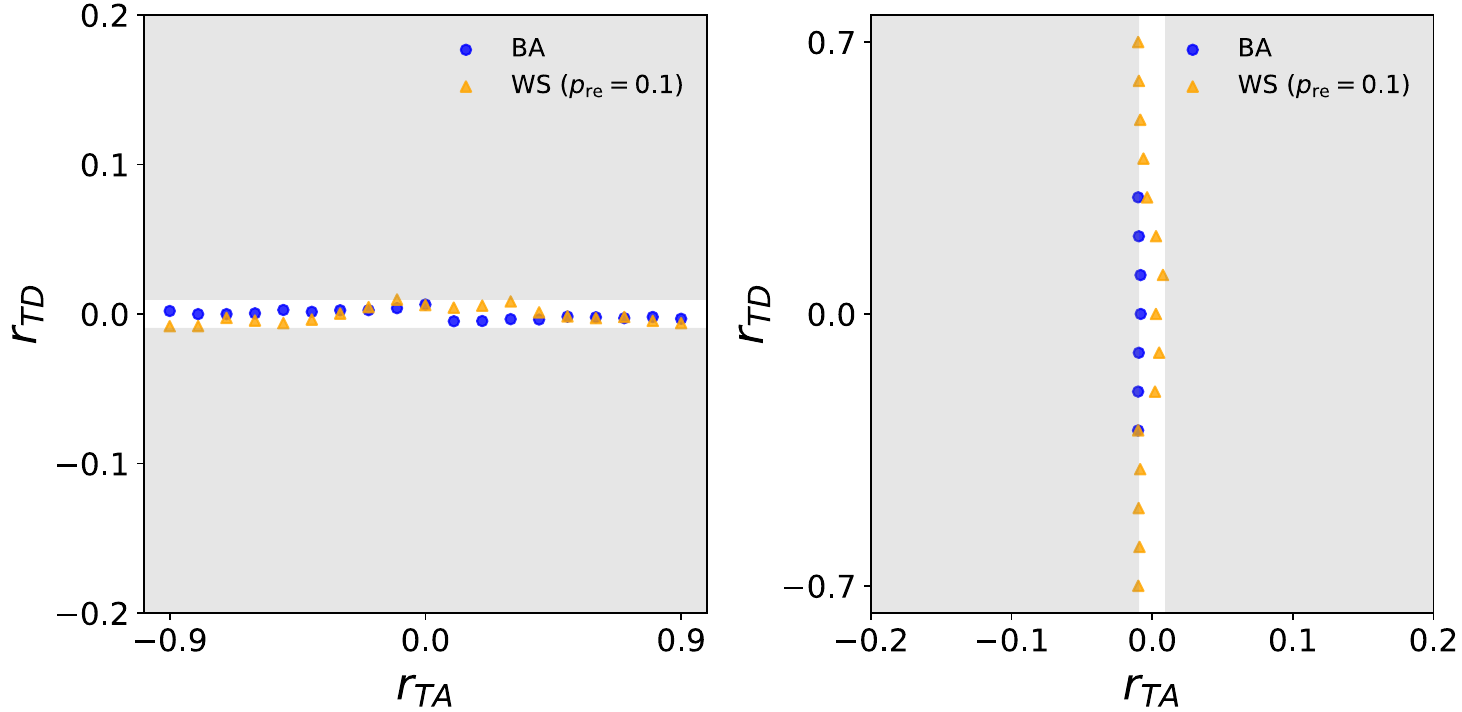}
    \caption{Quasi-pure samples for BA and WS networks with $N\times 100$ times of random pair swapping trials.}
    \label{fig-S5}
\end{figure*}

\begin{figure}[b]
    \includegraphics[width=\textwidth]{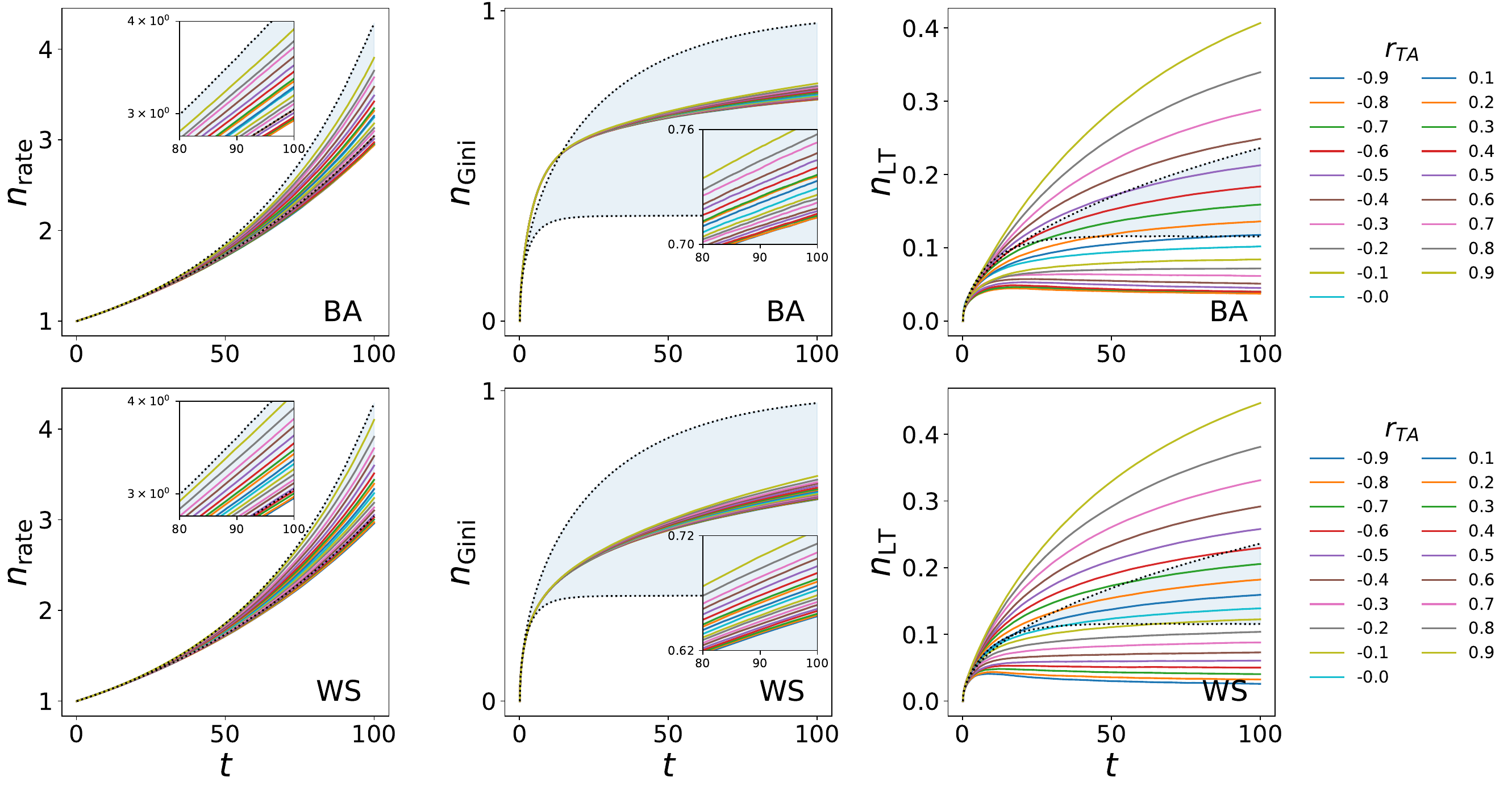}
    \caption{TA effect on time evolutions of three economic indices in the TLS model for BA and WS networks. The upper dashed line is the average of the TvL model, the lower dashed line is the average of the mean-field TLS model, and the light-blue shadow portion indicates the intermediate region between them. All simulations are performed for $(N,r,g,b,J)=(10^4,2,0.1,0.1,0.1)$ and $T_i\sim\mathcal{N}(0.6,0.1^2)$.}
    \label{fig-S6}
\end{figure}

\begin{figure}
    \includegraphics[width=\textwidth]{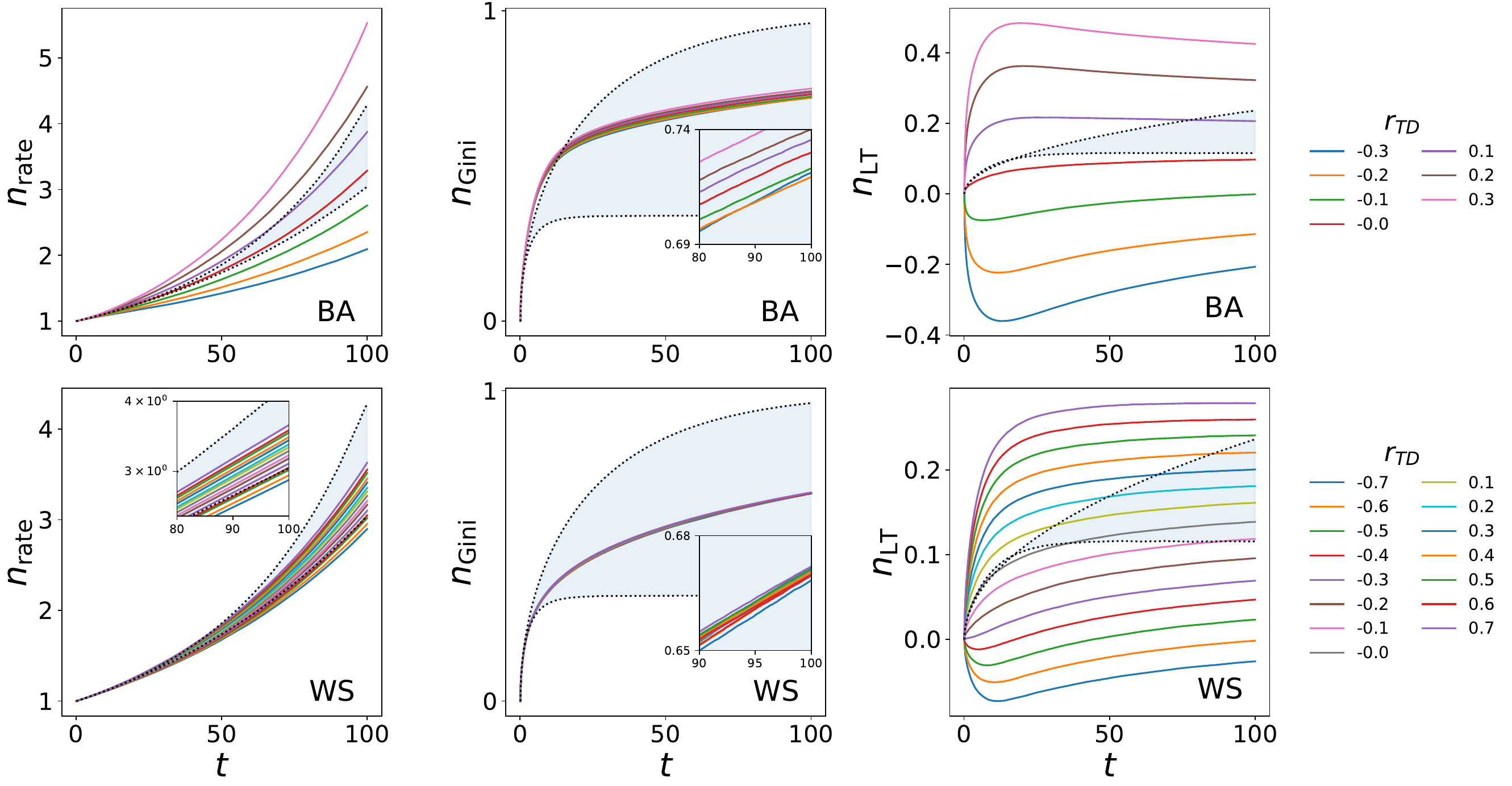}
    \caption{TD effect on time evolutions of three economic indices in TLS model for BA and WS networks. The upper dashed line is the average of the TvL model, the lower dashed line is the average of the mean-field TLS model, and the light-blue shadow portion indicates the intermediate region between them. All simulations are performed for $(N,r,g,b,J)=(10^4,2,0.1,0.1,0.1)$ and $T_i\sim\mathcal{N}(0.6,0.1^2)$.}
    \label{fig-S7}
\end{figure}

\subsection{\label{evolution} All indices evolution over quasi-pure $r_{\rm TA}$ and $r_{\rm TD}$ for BA and WS networks}
We show the time evolution of all three indices, where the parameter set $(N,r,g,b,J)=(10^4,2,0.1,0.1,0.1)$ and $T_i\sim\mathcal{N}(0.6,0.1^2)$. All lines are taken on the average indices of $2^{10}$ realizations. The upper dashed line is the average of the TvL model, the lower dashed line is the average of the mean-field TLS model, and the light-blue shadow portion indicates the intermediate region between them, see Fig.~\ref{fig-S6},~\ref{fig-S7}.

For $r_{\rm TD}\simeq0$, $n_{\rm rate}$ cannot be larger than the case of the TvL model, which is no matter how large $r_{\rm TA}$ value is. However, it can be smaller than case of the mean-field TLS model when $r_{\rm TA}$ is sufficiently small value. Thus, we suppose that $n_{\rm rate}$ of the TvL model is the upper bound of the TLS model without the talent-degree correlation effect ($r_{\rm TD}=0$) as the corresponding detailed hypothesis. Also, $n_{LT}$ cannot be larger than 0, which means the pure effect of $r_{\rm TA}$ cannot make a tendency for less talented agents to become richer.

Contrarily, for $r_{\rm TA}\simeq0$, $n_{\rm rate}$ can be larger than the case of the TvL model when $r_{\rm TD}$ is sufficiently large. It shows that choice and concentration to highly talented agent give some advantage to the system with the aspect of capital growth in the short-term regime. For the BA network with the `high-degree advantage', $n_{\rm Gini}$ almost lies in the intermediate range, only except in the short-term regime. The variances of $n_{\rm Gini}$ depend on $r_{\rm TA}$ more than $r_{\rm TD}$ for both cases.

\begin{figure*}[h]
 \includegraphics[width=\textwidth]{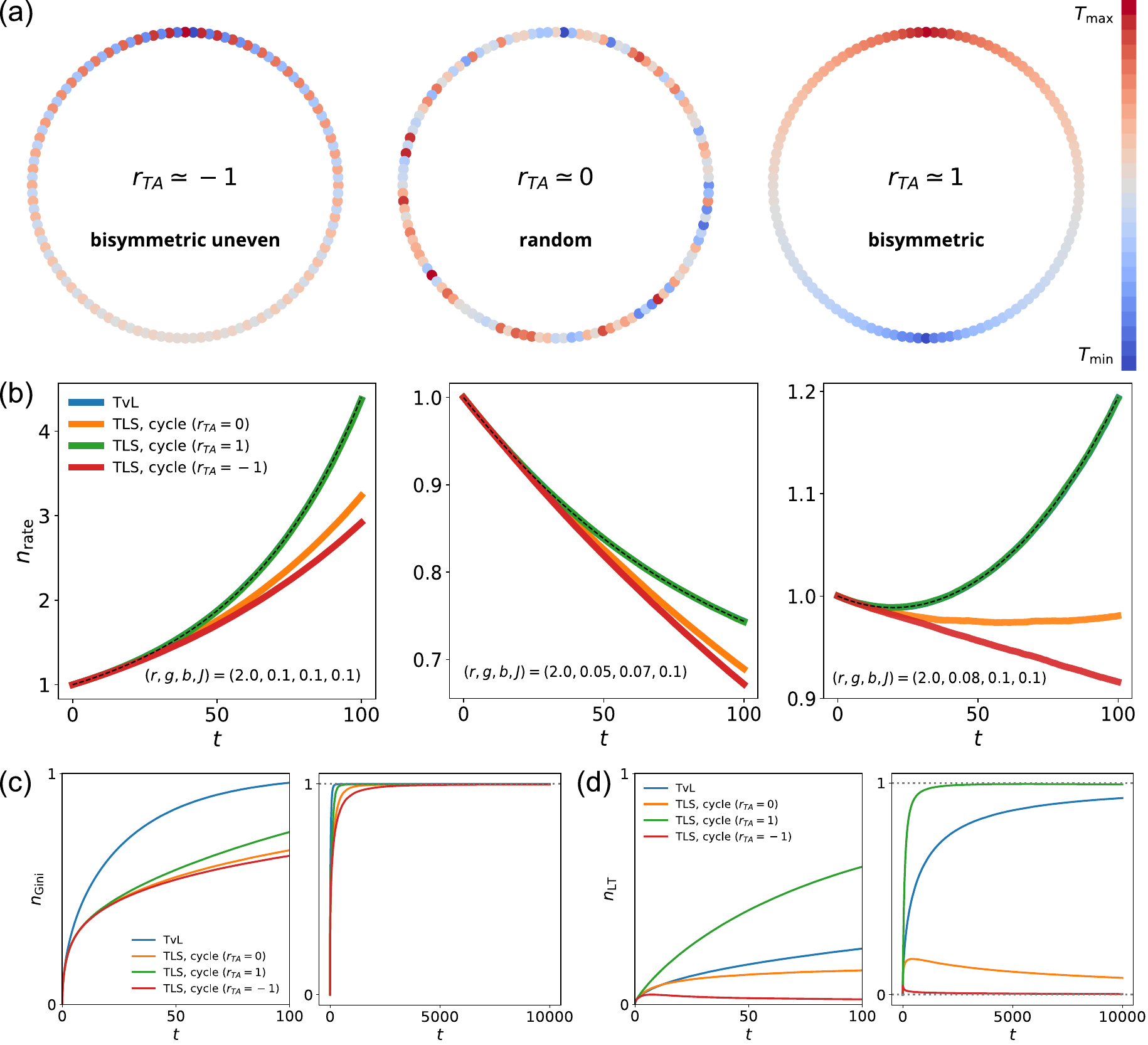}
    \caption{The TLS model on a cycle network: (a) Three representative talent configurations for a cycle network. (b) The time evolution of $n_{\rm rate}$ in the TLS model on a cycle network for various conditions of $(g,b)$. $(r,\mu,\sigma)=(2,0.6,0.1)$ used. (c) The time evolution of $n_{\rm Gini}$. (d) The time evolution of $n_{LT}$. $(r,g,b,\mu,\sigma)=(2,0.1,0.1,0.6,0.1)$ used for (c) and (d).}
    \label{fig-S8}
\end{figure*}

\section{\label{cycle}TLS model on cycle network}
We here discuss the role of talent configuration properties in the TLS model on the cycle network. To do so, we first define the talent assortativity $r_{\rm TA}$ as follows:
\begin{align}
    r_{\rm TA}=\frac{Cov(T,T')}{\sqrt{Var(T)Var(T')}}=\frac{\sum_{T}\sum_{T'}TT'(e_{TT'}-q_{T}q_{T'})}{\sum_{T}T^2q_{T}-(\sum_{T}Tq_{T})^2}=\frac{(1/M)\sum_{i}T_iT_i'-\left[(1/M)\sum_{i}\frac{T_i+T_i'}{2}\right]^2}{(1/M)\sum_{i}\frac{T_i^2+T_{i}'^2}{2}-\left[(1/M)\sum_{i}\frac{T_i+T_i'}{2}\right]^2},
    \label{eq-r_TA}
\end{align}
The last one in Eq.~\eqref{eq-r_TA} is the empirical representation of $r_{\rm TA}$, where $i$ is a link index, $T_i$ and $T'_i$ are nodal talents for the selected link $i$, and $M$ is the total number of links.

The cycle network can have the minimum and maximum of $r_{\rm TA}$ as -1 and 1, which correspond to the minimum and maximum values of Pearson correlation coefficient. Consider the infinitely large ($N\rightarrow\infty$) cycle network. If $T$ follows the normal distribution for our case, random variable $T$ can be represented by $T=\mu+\sigma Z$ ($Z\sim\mathcal{N}(0, 1^2)$). If talent configuration is bisymmetric, connected node talents are almost all same that means $T'\sim T$. For the bisymmetric uneven case, $T'\sim \mu-(T-\mu)$. Lastly, for the random case, connected talents can be represented by serial chains, like $(x_1y_1-x_2y_2-\cdots-x_Ny_N)$ and $(y_1x_2-y_2x_3-\cdots-y_Nx_1)$. Here $x_i$ and $y_i$ are the samples of two independent normal random variable $X$ and $Y$. For that case, all connected talent combinations are uncorrelated if $X$ and $Y$ are uncorrelated.
\begin{align}
    ({\rm bisymmetric\ uneven}) &: \quad r_{\rm TA}\simeq\frac{\langle{(T)(2\mu-T)}\rangle-\langle{T}\rangle\langle{2\mu-T}\rangle}{\sqrt{Var(T)Var(2\mu-T)}}=-\frac{\langle{T^2}\rangle-\langle{T}\rangle^2}{Var(T)}& =-1, \\
    ({\rm random}) &: \quad r_{\rm TA}\simeq\frac{\langle{XY}\rangle-\langle{X}\rangle\langle{Y}\rangle}{\sqrt{Var(X)Var(Y)}}=\frac{\langle{X}\rangle\langle{Y}\rangle-\langle{X}\rangle\langle{Y}\rangle}{\sqrt{Var(X)Var(Y)}}& =0, \\
    ({\rm bisymmetric}) &: \quad r_{\rm TA}\simeq\frac{\langle{TT}\rangle-\langle{T}\rangle\langle{T}\rangle}{\sqrt{Var(T)Var(T)}}=\frac{\langle{T^2}\rangle-\langle{T}\rangle^2}{Var(T)}& =1.
\end{align}

These results are consistent even if the talent distribution does not follow the normal distribution. When $N\rightarrow\infty$, if talent configuration is bisymmetric, the connected subset of nodes can be regarded as the long chain of agents sharing the same talent. If the edge effect can be neglected, the subset system can be considered as an isolated node represented by a subset size. This subset have mean capital as $$\langle{C}\rangle=C_0e^{\alpha(T)t}$$ (as the result of the BM model) and exist at the ratio of $\frac{1}{\sigma\sqrt{2\pi}}e^{-(T-\mu)^2/(2\sigma^2)}dT$ in the entire ring. Thus, in that case, the mean capital of the total system is $$\langle{C}\rangle=\frac{1}{\sigma\sqrt{2\pi}}\int_{-\infty}^{\infty}e^{-(T-\mu)^2/(2\sigma^2)}C_0e^{\alpha(T)t}dT,$$
exactly the same as the result of the TvL model. It is why the TLS model on the cycle network with $r_{\rm TA}=1$ have the same value of $n_{\rm rate}$ as that of the TvL model even in the different conditions of $(g,b)$, see Fig.~\ref{fig-S4}. (b). One more interesting thing is that for the case of $r_{\rm TA}=1$, $n_{LT}$ grows much faster than the case of the TvL model. This is quite an interesting phenomenon because capital should be transferred from those with more capital to those with less capital in a regular network. At a glance, it seems to disrupt meritocratic fairness because more capital on average comes from more talent in the long-term regime, and capital transfers are often from agents with more talent to agents with less talent. For $r_{\rm TA}\neq 1$, it is true that the redistributive interaction of the TLS model in the cycle network disrupts meritocratic fairness since $n_{LT}$ is smaller than that of the TvL model (has no interaction) in the long-term regime. However, for $r_{\rm TA}=1$, interaction leads the system to a perfectly meritocratic society much faster than the TvL model, see Fig.~\ref{fig-S8} (d). In this case, it rather promotes the system's meritocratic fairness, and it can be considered as ``cartel effect" mentioned in the main text. On the other hand, for the case of $r_{\rm TA}=-1$, $n_{LT}$ converges to 0 after a long time which means there is no correlation between capital level $L$ and talent $T$.

\end{widetext}
\end{document}